%% file: hybrid22.tex
\documentclass[11pt, letter]{article}

\usepackage{stolenstyle}

\usepackage{amsmath,epsf,amssymb,latexsym,amsthm,setspace,bbm,array,pifont}

\usepackage[matrix,arrow,frame,import,curve,color]{xy}

\input{basedefs2}

\def\br{{\boldsymbol{r}}}
\def\bx{{\boldsymbol{x}}}

\def\bK{{{\boldsymbol{K}}}}

\def\bY{{{\boldsymbol{Y}}}}
\def\bYk{{{\bY_{\!\!k}}}}

\def\cXb{{{\overline{\cX}}}}
\def\cYb{{{\overline{\cY}}}}

\def\bQ{{{{\boldsymbol{Q}}}}}
\def\bQb{{{\overline{\bQ}}}}

\def\bq{{{{\boldsymbol{q}}}}}
\def\bqb{{{\overline{\bq}}}}

\def\pbz{{{\pb_{\!\zb}}}}
\def\pz{{{\p_{\!z}}}}
\def\Dbz{{{\Db_{\!\zb}}}}
\def\Dz{{{D_{\!z}}}}

\def\pp#1{{{\frac{\p}{\p#1}}}}

\def\vec#1{{{\bf#1}}}

\def\tkin{{{\text{kin}}}}

\def\LL{{\mathbb L}}

\def\Sym{\operatorname{Sym}}
\def\obs{{\operatorname{obs}}}

\def\preprint{ {{AEI-2013-230}}}

\title{Hybrid conformal field theories}
\author[a] {Marco Bertolini,}
\author[b] {Ilarion V.~Melnikov,}
\author[a] {and M.~Ronen Plesser}
\affiliation[a]{Center for Geometry and Theoretical Physics, Box 90318 \\
Duke University, Durham NC 27708-0318, USA}
\affiliation[b]{Max-Planck-Institut f\"ur Gravitationsphysik (Albert-Einstein-Institut)\\
 Am M\"uhlenberg 1, D-14476 Golm, Germany}
\emailAdd{mb266@phy.duke.edu}
\emailAdd{ilarion@aei.mpg.de}
\emailAdd{plesser@cgtp.duke.edu}
\abstract{We describe a class of (2,2) superconformal field theories obtained by fibering a Landau-Ginzburg orbifold CFT over a compact K\"ahler base manifold.  While such models are naturally obtained as phases in a gauged linear sigma model, our construction is independent of such an embedding.  We discuss the general properties of such theories and present a technique to study the massless spectrum of the associated heterotic compactification.  We test the validity of our method by applying it to hybrid phases of linear models and comparing spectra among the phases.}

\begin{document}

\maketitle

\section{Introduction}\label{s:intro}

Just what is a hybrid anyway?  In constructing two-dimensional superconformal field theories (SCFTs) relevant for superstring vacua
we are used to two sorts of massless fluctuating fields:  those corresponding to a non-linear sigma model (NLSM), and those corresponding to a Landau-Ginzburg (LG) theory.  The former define a classically conformally invariant system.  Under favorable conditions, e.g. a Calabi-Yau target space and world-sheet supersymmetry, the background fields can be chosen to preserve superconformal invariance, and when the background is weakly coupled in a ``large radius limit'' (i.e. the background fields have small gradients), the theory reduces to a free-field limit. The latter have superpotential interactions that explicitly break scale invariance; however, under favorable conditions, e.g. a quasi-homogeneous superpotential, the IR limit of such a theory defines a non-trivial SCFT.   

In each case, the utility of the description is two-fold:  at a fundamental level, we can use the weakly coupled UV theory to define a SCFT; as a practical matter, the weakly coupled description, combined with non-renormalization theorems that follow from supersymmetry, allow us to identify and compute certain protected quantities such as chiral rings and massless spectra of the associated string vacua in terms of the UV degrees of freedom.  

By now the reader has surely guessed what is meant by a hybrid~\cite{Witten:1993yc,Aspinwall:1993nu}:  it is a two-dimensional theory that includes both types of massless fluctuating fields: ones that have classically conformally invariant NLSM self-interactions, as well as some that self-interact via a superpotential; of course an interesting hybrid also has interactions between the two types of degrees of freedom.  A hybrid is a fibered theory, where the fiber is a LG theory with potential whose coefficients depend on the fields of the base NLSM.  The potential is chosen so that its critical point set is the base target space.  
We then have two important questions:  what are the criteria for a hybrid theory to flow to a SCFT?  how do we generalize NLSM/LG techniques to compute physical quantities?

It is well-known that all of these descriptions--- large radius limits of NLSMs, Landau-Ginzburg orbifolds (LGOs), and hybrid loci---arise as phases of (2,2), and more generally (0,2) gauged linear sigma models (GLSMs)~\cite{Witten:1993yc}.  The GLSM philosophy is that each phase should yield a limiting locus where at least protected quantities should be amenable to computation via the UV weakly-coupled field theory description.  Such techniques are known for large radius NLSM and LGO phases but not for more general phases.  In this work, we take a step in developing techniques for what we will call the ``good hybrid'' phases of a GLSM.\footnote{Along the way we obtain a simple and direct description of the massless spectrum for the large radius limit of a (0,2) NLSM --- an application to CY NLSMs with non-standard embedding may be found in appendix~\ref{app:02NLSM}~.}  

Although this does not cover a generic GLSM phase, and there are perhaps good reasons~\cite{Aspinwall:2009qy} that we should not expect a simple description for a generic phase, it does increase the set of special points in the moduli space amenable to exact computations; this can lead to useful insights into stringy moduli space as in~\cite{Aspinwall:2010ve,Aspinwall:2011us,Aspinwall:2011vp,Blumenhagen:2011sq}.  In addition, our definition of a good hybrid model, although inspired by the GLSM construction, will not explicitly invoke the GLSM.~ Thus, we are in principle providing a new class of UV theories that can lead to SCFTs without a known GLSM embedding.  

In this note we will focus on hybrid theories with (2,2) world-sheet supersymmetry that are suitable for supersymmetric string compactification, i.e. ones with integral  $\GUL\times\GUR$ R-symmetry charges; as in the case of LGO string vacua, this is achieved by taking an appropriate orbifold.

While such models offer a good point of departure, it is clear that a more general (0,2) hybrid framework will be both computationally useful and conceptually illuminating.  We will describe (0,2) hybrids in a future work; for now we note that just like (2,2) LG models, the hybrids incorporate a class of Lagrangian deformations away from the (2,2) locus.  These are obtained by smoothly deforming the (2,2) superpotential to a more general (0,2) form. 

In what follows, we first give a broad geometric description of (2,2) hybrids, construct a Lagrangian for a good hybrid model and study its symmetries.  With that basic structure in hand, we turn to a technique, valid in the large base volume limit and generalizing the well-known (2,2) and (0,2) LGO results of~\cite{Kachru:1993pg,Distler:1993mk}, to compute the massless heterotic spectrum of a hybrid compactification.  We then apply the techniques to a number of examples and conclude with a brief discussion of applications and further directions.  

\acknowledgments  It is a pleasure to thank N.~Addington, P.S.~Aspinwall, A.~Degeratu, S.~Katz, D.R.~Morrison, and E.~Sharpe for useful discussions.
IVM is grateful to University of Alberta and BIRS for hospitality while this work was being completed.  MRP thanks the Aspen Center for Physics and NSF Grant 1066293 and the University of California at Santa Barbara for hospitality during work on this project, as well as the members of the Skywackers club for their hospitality at Larry Dennis flight park where some of this work was completed.
MB thanks the Max-Planck-Institut f\"ur Gravitationsphysik (Golm) for hospitality during the completion of this work.
MB and MRP are supported by NSF Grant PHY-1217109.

\section{A geometric perspective} \label{s:geom}

The geometric setting for our theory is a (2,2) NLSM constructed with (2,2) chiral superfields.  Consider a K\"ahler manifold $\bY_{\!\!0}$ equipped with a holomorphic function---the superpotential $W$---chosen so that its critical point set is a compact subset $B\subset \bY_0$.  More precisely, $dW$, a holomorphic section of the cotangent bundle $T_{\bY_{\!\!0}}^\ast$, has the property that $dW^{-1}(0) = B \subset \bY_{\!\!0}$.  We call this \emph{the potential condition}.  A LG model, with $\bY_{\!\!0} \simeq \C^n$ and $B$ being the origin, is a familiar example.  A compact $\bY_{\!\!0}$ necessarily has a trivial superpotential, and the resulting theory is just a standard compact NLSM.

We say a geometry satisfying the potential condition has a \emph{hybrid model} iff the local geometry for $B\subset \bY_{\!\!0}$ can be modeled  by $\bY$ --- the total space of a rank $n$ holomorphic vector bundle $X \to B$ over a compact smooth K\"ahler base $B$ of complex dimension $d$.  The point of this definition is that the superpotential interactions will lead to a suppression of finite fluctuations of fields away from $B$, so that the low energy physics of the original NLSM will be well-approximated by the restriction to the hybrid model.  Our main task will be to describe this low energy physics, and in what follows we will concentrate on the hybrid model geometry $\bY$.  In many examples (e.g. the LG theories) $\bY \simeq \bY_{\!\! 0}$, but our results apply to the more general situation where $\bY$ is simply a local model.
 A simple example of a hybrid geometry, where $X = \cO(-2)$ over $B = \P^1$, is presented in appendix~\ref{app:simplehybrid}.

In order to be reasonably confident that the low energy limit of a hybrid model is a (2,2) SCFT, we will need the geometry to satisfy several additional conditions intimately related to the existence of chiral symmetries and GSO projections.  It will be easiest to discuss these after we introduce the explicit Lagrangian realization of this geometry.  In
our examples these features will already be present in the ``UV'' completion of the hybrid model, offered either by $\bY_{\!\!0}$ or some other high energy description such as a GLSM.\footnote{It would be interesting to find hybrid examples where these features emerge accidentally.}

A final geometric comment, relevant for heterotic applications, concerns (0,2)-preserving deformations of these theories.
(2,2) theories often admit a class of smooth (0,2) deformations, where the left-moving fermions couple to $\cE\to \bY$, a deformation of $T_{\bY}$, and the (0,2) superpotential is encoded by a holomorphic section $J \in \Gamma(\cE^\ast)$ with $J^{-1}(0) = B$.  In the hybrid case there exist (0,2) deformations where $\cE = T_{\bY}$ but $d J \neq 0$; such a (0,2) superpotential cannot be integrated to a (2,2) superpotential $W$.  Turning these on leads to a simple class of (0,2) hybrid models.  

\section{Action and symmetries} \label{s:action}
In this section we construct the (2,2) SUSY UV action for a hybrid model and analyze its symmetries.
We begin with the necessary superspace formalism for a flat Euclidean world-sheet with coordinates $(z,\zb)$.
Since we are interested in (0,2) deformations of (2,2) theories, it will be convenient for us to work with both (2,2)
and (0,2) superspaces.\footnote{Our superspace conventions are those of~\cite{Melnikov:2011ez}; more details may be found in~\cite{Distler:1995mi} or~\cite{West:1990tg}.}  Let's start with the latter.  Introducing Grassmann coordinates $\theta$ and $\thetab$, we
obtain the supercharges
\begin{align}
\cQ = -\frac{\p}{\p\theta} + \thetab \pbz,\qquad
\cQb = -\frac{\p}{\p\thetab} +\theta \pbz,
\end{align}
where $\pb_{\!\zb} \equiv \p/\p \zb$.  These form a representation of the (0,2) SUSY algebra:  $\cQ^2 = \cQb^2 =  0$ and $\AC{\cQ}{\cQb} = -2\pbz$.  The supercharges are graded by a $\GUR$ symmetry that assigns charge $\bqb = 1$ to $\theta$, and they anticommute with the supercovariant derivatives
\begin{align}
\cD = \frac{\p}{\p\theta} + \thetab \pbz,\qquad
\cDb = \frac{\p}{\p\thetab} +\theta \pbz,
\end{align}
that satisfy $\cD^2 =\cDb^2 = 0$ and $\AC{\cD}{\cDb} = 2\pbz$.  

To build a (2,2) superspace we introduce additional Grassmann variables $\theta',\thetab'$ and form $\cQ'$, $\cQb'$, as well as $\cD'$ and $\cDb'$, by replacing $(\theta,\thetab,\pbz) \to (\theta',\thetab',\pz)$, where $\pz = \p /\p z$.  These supercharges and derivatives are graded by $\GUL$ that assigns charge $\bq=1$ to $\theta'$.

\subsection{Multiplets}
We are interested in K\"ahler hybrid models with target space $\bY$, and these can be constructed by using bosonic chiral (2,2) superfields and their conjugate anti-chiral multiplets\footnote{Recall that a chiral superfield $A$ satisfies the constraints $\cDb A = \cDb' A = 0$; more general (2,2) multiplets (twisted chiral and semi-chiral) are reviewed in, for instance,~\cite{Lindstrom:2005zr}.} denoted by $\cY^\alpha$ and $\cYb^{\alphab}$, with $\alpha,\alphab = 1,\ldots, \dim \bY$.  These decompose into (0,2) chiral and anti-chiral multiplets as follows:
\begin{align}
\label{eq:02sfields}
\cY^\alpha &= Y^\alpha + \sqrt{2} \theta' \cX^\alpha + \theta'\thetab' \pz Y^\alpha~, &&&
\cYb^{\alphab} & = \Yb^{\alphab} - \sqrt{2}\thetab' \cXb^{\alphab} -\theta'\thetab' \pz \Yb^{\alphab}~,\nonumber\\[2mm]
Y^\alpha & = y^\alpha + \sqrt{2}\theta \eta^\alpha +\theta\thetab \pbz y^{\alpha}~,&&&
\Yb^{\alphab} & = {\yb}^{\alphab} - \sqrt{2}\thetab \etab^{\alphab} -\theta\thetab \pbz \yb^{\alphab}~, \nonumber\\
\cX^\alpha & = \chi^\alpha + \sqrt{2}\theta H^\alpha +\theta\thetab \pbz \chi^{\alpha}~,&&&
\cXb^{\alphab} & = {\chib}^{\alphab} + \sqrt{2}\thetab \Hb^{\alphab} -\theta\thetab \pbz \chib^{\alphab}~.
\end{align}
The $Y^\alpha$ are bosonic (0,2) chiral multiplets, while the $\cX^\alpha$ are chiral fermi multiplets, with lowest component a left-moving fermion $\chi^\alpha$; the $H^\alpha$ and their conjugates are auxiliary non-propagating fields.\footnote{A comment on Euclidean conventions:  the charge conjugation operator $\cC$, inherited from Minkowski signature, conjugates the complex bosons and acts as $\cC(\chi) = \chib$ and $\cC(\chib) =-\chi$ for every fermion $\chi$.}

Since $\bY$ is the total space of a vector bundle, it will occasionally be useful to split the $y^\alpha$ into base and fiber coordinates, which we will denote by $y^\alpha = (y^I, \phi^i)$, with $I = 1,\ldots, d$ and $i = 1,\ldots, n$.  The $y^I$ are then coordinates on the base manifold $B$, while the $\phi^i$ parametrize the fiber directions.  

\subsection{The (2,2) hybrid action}
The two-derivative (2,2) action is a sum of kinetic and potential terms, with 
\begin{align}
\label{eq:22action}
S_{\tkin} &= \frac{1}{4\pi} \int d^2 z ~\cD \cDb \cL_{\tkin},\qquad \cL_{\tkin} = \frac{1}{2} \cDb' \cD' \bK(\cY,\cYb),\nonumber\\
S_{\text{pot}} &= \frac{\sqrt{2}m}{4\pi} \int d^2z ~\cD \cW(Y,\cX) + \text{c.c.},\qquad
 \cW =\frac{1}{\sqrt{2}} \cD' W(\cY)~.
\end{align}
As is well-known, the kinetic term leads to a $\bY$ NLSM with a K\"ahler metric $g$.
The superpotential $W$ is a holomorphic function on $\bY$ satisfying the potential condition, i.e. $dW(0)^{-1} = B$; $m$ is a parameter with dimensions of mass.  If the metric $g$ is well-behaved, then the potential condition leads a suppression of field fluctuations away from $B \subset \bY$ via the bosonic potential
\begin{align}
\label{eq:bospot}
S \supset \frac{|m|^2}{2\pi} \int d^2 z ~g^{\alpha\betab} \p_\alpha W \p_{\betab} \Wb~,
\end{align}
and at low energies (as compared to $|m|$) the kinetic term can be taken to be quadratic in the fiber directions, i.e. the K\"ahler potential is
\begin{align}
\label{eq:Kbf}
\bK= K(y^I,\yb^{\Ib}) + \phi h(y^I,\yb^{\Ib}) \phib + \ldots,
\end{align}
where $K$ is a K\"ahler potential for a metric on $B$, $h$ is a Hermitian metric on $X \to B$, and $\ldots$ denotes neglected terms in the fiber coordinates.  Using the base--fiber decomposition the metric $g_{\alpha\betab} = \p_\alpha \p_{\betab} \bK \equiv \bK_{\!\alpha\betab} $ then takes the form
\begin{align}
g = (K_{I\Jb} - \phi \cF_{I\Jb} h \phib) dy^I d\yb^{\Jb} + D\phi h \Db\phib + \ldots,
\end{align}
where  $\cA = \p h h^{-1}$ is the Chern connection for the metric $h$, $\cF = \pb \cA$ is its (1,1) curvature, and 
$D\phi = d\phi +\phi \cA$ is the corresponding covariant derivative.

\subsubsection*{Positivity of the metric and the case $\bY \simeq \bY_{\!\! 0}$}
In many cases we need not worry about higher order corrections to $g$ in order to define a sensible theory.  As in the simple case of LG models, this would be a situation where we need not consider the distinction between $\bY$ and $\bY_{\!\! 0}$ from above.  Examining the form of $g$, we see that a necessary condition is that $\phi \cF_{I\Jb} h \phib$ is non-positive for all points in $\bY$.\footnote{Suppose there is a point $p\in B$ and $\phi_0 \in \pi^{-1}(p)$ such that the Hermitian form $\phi_0 \cF_{I\Jb} h \phib_0$ has a positive eigenvalue.  Then taking $\phi = t \phi_0$, for sufficiently large $t$ the metric $g$ will cease to be positive.}  We say a bundle $X\to B$ is non-positive if it admits a Hermitian metric $h$ that satisfies this non-positivity condition.

Thus, to use~(\ref{eq:Kbf}) to define a UV-complete theory, we are led to a geometric question:  what are the non-positive bundles over $B$?  This is closely related to classical questions in algebraic geometry regarding positive and/or ample bundles, and using those classical results we can easily find sufficient conditions for non-positivity.  Recall that a line bundle $L\to B$ is said to be positive if its (1,1) curvature form is positive; it is said to be negative if the dual bundle $L^\ast$ is positive~\cite{Griffiths:1978pa,Lazarsfeld:2004pa}.  Taking $X = \oplus_i L_i$,  a sum of negative and trivial line bundles, leads to many examples of non-positive bundles.   

We should stress two points:  first, even this set of examples leads to many previously unexplored SCFTs.  Second, more generally, we do not need to assume that $\bY \simeq \bY_{\!\!0}$ or that $g$ has no higher-order terms in the fibers.  The low energy limit of a UV theory with a hybrid model will be well-described by our action, and the potential condition will imply that the fiber corrections to the metric will not be important to the low energy physics.   We will analyze one such example below, where $X$ is a sum of a positive and a negative bundle.

\subsubsection*{(0,2) action}
Since we are interested in heterotic applications as well as (0,2) deformations, it is useful to have the manifestly (0,2) supersymmetric action obtained by  integrating over $\theta',\thetab'$ in~(\ref{eq:22action}).  Absorbing the superpotential mass scale $m$ into $W$ the result is
\begin{align}
\label{eq:02action}
\cL_{\tkin} &= \ff{1}{2} (\bK_\alpha \pz Y^\alpha- \bK_{\alphab} \pz \Yb^{\alphab}) + g_{\alpha\betab} \cX^\alpha \cXb^{\alphab}~, &
\cW & = \cX^\alpha W_\alpha~.
\end{align}
where $\bK_{\!\alpha} \equiv \p \bK /\p Y^\alpha$, $W_\alpha \equiv \p W/\p Y^\alpha$, etc. 
It is a simple matter to obtain the classical equations of motion from the (0,2) action.\footnote{If $A$ and $B$ are (0,2) superfields, then $\cD\cDb(A B) |_{\theta,\thetab =0} = 0 \quad \forall B \implies A = 0$;  any chiral (anti-chiral) superfield, say $\delta \cX$ ($\delta \cXb$), can be expressed as $\cDb P$ ($\cD \Pb$) for some superfield $P$.} 
The result is
\begin{align}
\label{eq:02eom}
\cDb~\cXb_{\alpha} = \sqrt{2} W_\alpha,\qquad
\cDb\left[ g_{\alpha\betab} \p \Yb^{\betab} +g_{\alpha\betab,\gamma} \cXb^{\betab}\cX^\gamma  \right]  = \sqrt{2} \cX^\beta W_{\alpha\beta}~,
\end{align}
where we defined the fermi superfield $\cXb_{\alpha} \equiv g_{\alpha\betab}(Y,\Yb) \cXb^{\betab}$. 

\subsubsection*{Component action}
Finally, we can integrate over the remaining (0,2) superspace coordinates $\theta$ and $\thetab$ to obtain the component action.  The auxiliary field $\Hb^{\alphab}$ is determined by the equations of motion~(\ref{eq:02eom}):
\begin{align}
 g_{\alpha\betab}\Hb^{\betab} = g_{\alpha\betab,\gammab}\etab^{\gammab}\chib^{\betab} + W_\alpha~,
\end{align}
and using this as well as $\chib_\alpha \equiv g_{\alpha\betab} \chib^{\betab}$ we obtain
\begin{align}
\label{eq:fullcomponent}
2\pi L & = g_{\alpha\betab} \left( \pbz y^\alpha\pz\yb^{\betab} + \etab^{\betab} \Dz \eta^\alpha\right) + \chib_{\alpha} \Dbz \chi^\alpha -\etab^{\betab} \eta^\alpha R_{\alpha\betab\gamma}^{~~~~\delta} \chib_{\delta} \chi^\gamma
- \chi^\alpha \eta^\beta D_\beta W_{\alpha} \nonumber\\
&\qquad+\chib^{\alphab}\etab^{\betab} D_{\betab}\Wb_{\alphab} + g^{\betab\alpha} W_\alpha \Wb_{\betab}~,
\end{align}
where the covariant derivatives are defined with the K\"ahler connection $\Gamma^\alpha_{\beta\gamma} \equiv g_{\gamma\betab,\beta} g^{\betab \alpha}$, e.g.
\begin{align}
\Db_{\zb} \chi^\alpha = \pbz \chi^\alpha +  \pbz y^{\betab} \Gamma^\alpha_{\beta\gamma} \chi^\gamma~,\qquad
D_\alpha W_\beta = W_{\beta\alpha} - \Gamma^\gamma_{\alpha\beta} W_\gamma~,
\end{align}
and the curvature is $R_{\alpha\betab\gamma}^{~~~~\delta} \equiv \Gamma^\delta_{\alpha\gamma,\betab}$.
This is a complicated interacting theory, and in general it is not clear that one set of fields is preferred to another (say using $\chib_\alpha$ instead of $\chib^{\alphab}$); however, for the purpose of determining the massless spectrum, it turns out to be useful to introduce another field redefinition to keep track of the non-zero left-moving bosonic excitations:
\begin{align}
\label{eq:rhodef}
\rho_\alpha \equiv g_{\alpha\alphab} \p \yb^{\alphab} + \Gamma^\delta_{\alpha\gamma}\chib_\delta \chi^\gamma~,
\end{align}
in terms of which the left-moving kinetic terms take a strikingly simple form:
\begin{align}
\label{eq:rhoaction}
2\pi L & = \rho_\alpha \pbz y^\alpha + \chib_\alpha \pbz \chi^\alpha + \eta^\alpha\left[ g_{\alpha\betab} \Dz \etab^{\betab} + \etab^{\betab} R_{\alpha\betab\gamma}^{~~~~\delta} \chib_\delta\chi^\gamma +\chi^\beta D_\alpha W_\beta
\right]~
\nonumber\\
&\qquad+\chib^{\alphab}\etab^{\betab} D_{\betab}\Wb_{\alphab} + g^{\betab\alpha} W_\alpha \Wb_{\betab}~.
\end{align}
Unlike the other fields $\rho$ does not transform as a section of the pull-back of a bundle on $\bY$ under target space diffeomorphisms; this will have important consequences below.

\subsection{Symmetries}
We now examine the symmetries of the hybrid Lagrangian.

\subsubsection*{The $\bQb$ supercharge}
Our action respects (2,2) SUSY generated by the superspace operators $\cQ$ and $\cQb$, as well as their left-moving images.  We define the action of the corresponding operators $\bQ$ and $\bQb$ by
\begin{align}
\sqrt{2} \CO{\xi \bQ + \xib \bQb}{A} \equiv - \xi \cQ A -\xib \cQb A,
\end{align}
where $\xi$ is an anti-commuting parameter and $A$ is any superfield.  In order to avoid writing the graded commutator, we will use a condensed notation $\xib \bQb \cdot A \equiv \CO{\xib \bQb}{A}$.  For our subsequent study of the right-moving Ramond ground states, we will be particularly interested in the action of $\bQb$.  Using the superfields in~(\ref{eq:02sfields}),  we obtain
\begin{align}
\label{eq:Qbarfull}
\bQb \cdot y^\alpha & = 0, & 
\bQb \cdot \chi^\alpha & = 0, &
\bQb \cdot \eta^\alpha &= \pbz y^\alpha~, &
\bQb \cdot H^\alpha  & =  \pbz \chi^\alpha~, 
\nonumber\\
\bQb \cdot \yb^{\alphab} & =-\etab^{\alphab}~, &
\bQb \cdot \chib^{\alphab}& = \Hb^{\alphab}~, &
\bQb \cdot \etab^{\alphab} & = 0~,& 
\bQb \cdot \Hb^{\alphab} & = 0~.
\end{align}
The action of the remaining supercharges is easily obtained from this one by conjugation and/or switching left- and right-moving fermions.  Eliminating the auxiliary fields by their equations of motion we obtain
\begin{align}
\label{eq:QbarHeom}
\bQb \cdot y^\alpha & = 0~,&
\bQb \cdot \chi^\alpha  &= 0~,&
\bQb \cdot \eta^\alpha &= \pbz y^\alpha~, \nonumber\\
\bQb \cdot \yb^{\alphab} & = -\etab^{\alphab}~,&
\bQb \cdot \chib_\alpha  &= W_\alpha~, &
\bQb \cdot \etab^{\alphab}  &= 0~.
\end{align}
From~(\ref{eq:02eom}) it follows that up to the $\etab$ equations of motion we also have $\bQb\cdot \rho_\alpha = \chi^{\beta} W_{\beta\alpha}$.
Hence we can decompose  $\bQb$ as $\bQb = \bQb_0 + \bQb_W$, where the non-trivial action is
\begin{align}
\label{eq:supersplit}
\bQb_0 \cdot \yb^{\alphab} &= -\etab^{\alphab}~,&
\bQb_0 \cdot \eta^\alpha &= \pbz y^\alpha~,&
\bQb_W \cdot \chib_\alpha & = W_\alpha~,&
\bQb_W \cdot \rho_\alpha & = \chi^\beta  W_{\beta\alpha}~. 
\end{align}
These satisfy $\bQb_0^2 = \bQb_W^2 = \AC{\bQb_0}{\bQb_W} = 0$.\footnote{If we keep the terms in $\bQb \cdot \rho$ proportional to $\etab$ equations of motion and decompose that into a $W$-independent and $W$-dependent contributions, we find that the decomposition $\bQb = \bQb_0 + \bQb_W$ into a pair of nilpotent anti-commuting operators holds without use of equations of motion; for us the result of~(\ref{eq:supersplit}) will be sufficient.}  $\bQb_0$ is the supercharge for the NLSM with $W=0$, while $\bQb_W$ incorporates the effect of a non-trivial potential.

\subsubsection*{Chiral $\GU(1)$ symmetries}
The $\GUL\times\GUR$ symmetries play an important role in relating the UV hybrid model to the IR physics of the corresponding SCFT.\,   In the classical NLSM with $W=0$ the presence of these symmetries is a consequence of the existence of an integrable, metric-compatible complex structure on $\bY$.
In terms of components fields, the symmetries leave the bosonic fields invariant, while rotating the fermions as follows:  
\begin{align}
\GUL^0 ~:~ & \delta_{L}^0 \eta = 0,\qquad \delta^0_{L} \chi =- i \ep \chi~; & 
\GUR^0 ~:~ & \delta_{R}^0 \eta = -i\ep \eta,\qquad \delta^0_{R} \chi = 0~,
\end{align}
where $\ep$ is an infinitesimal real parameter.
These naive symmetries are explicitly broken by the superpotential, but they can be improved if the geometry $(\bY,g)$ admits a holomorphic Killing vector $V$ satisfying $\cL_{V} W = W$.\footnote{Holomorphic Killing vectors satisfy $V^\alpha_{,\betab} = 0$ and $\cL_V g = 0$.  They are a familiar topic in supersymmetry---see, e.g.,  Appendix D of~\cite{Wess:1992cp}.  Note that on a compact K\"ahler manifold a Killing vector field is holomorphic, but this can fail on a non-compact manifold.  Killing vectors on K\"ahler manifolds are further discussed in~\cite{Ballmann:2006le,Moroianu:2007fk}.}  $V$ generates a non-chiral symmetry action
\begin{align}
\delta_{V} Y^\alpha = i\ep V^\alpha(Y),\quad
\delta_{V} \Yb^{\alphab} = -i \ep \Vb^{\alphab}(\Yb);\qquad
\delta_{V} \cX^\alpha = i\ep V^\alpha_{,\beta} \cX^\beta~,
\delta_{V} \cXb^{\alphab} = -i \ep \Vb^{\alphab}_{,\betab} \cXb^{\betab}~,
\end{align}
and it is easy to see that $\delta_{L,R}\equiv \delta_{L,R}^0 + \delta_V$ are symmetries of the classical action.

While $\GU(1)_{\text{diag}} \subset \GUL\times\GUR$ has a non-chiral action on the fermions and hence is non-anomalous, $\GUL$ is a chiral symmetry that will be anomaly free iff $c_1(T_{\bY}) = 0$, a condition satisfied when $\bY$ is a non-compact Calabi-Yau manifold, i.e. $\bY$ has a trivial canonical bundle $K_{\bY} \simeq \cO_{\bY}$.  In what follows we assume $K_\bY$ is indeed trivial (this is stronger than $c_1(T_{\bY}) = 0$). 
When $X = \oplus_i L_i$, a sum of line bundles such that $\otimes_i L_i$ is negative, then since $K_{\bY} = K_B \otimes_i L_i^\ast$ the anti-canonical class of $B$ is very ample and $B$ is Fano.\footnote{A variety is Fano iff its anti-canonical class is ample; Fano varieties are quite special: for instance $H^i(B,\cO) =0$ for $i>0$, $\Pic(B)\simeq H^2(B,\Z) $; in addition, they are classified in dimension $d \le 3$ and admit powerful criteria for evaluating positivity of bundles~\cite{Lazarsfeld:2004pa}.}

In what follows we will denote the conserved charge for $\GUL$ ($\GUR$) by $J_0$ ($\Jb_0$) and its eigenvalues on various operators and states by $\bq$ ($\bqb$).

\subsubsection*{R-symmetries for good hybrid models}
We would like to identify the UV $\GUL\times\GUR$ symmetries described above with their counterparts in the conjectured IR SCFT.  As usual, there is a small subtlety in doing this when $V$ is not unique.  In practice this is easily achieved by picking a sufficiently generic superpotential and more generally, one could use $c$-extremization~\cite{Benini:2012cz} to fix $\GUL\times\GUR$ up to the usual caveats of accidental IR symmetries.

More importantly, in order for the UV R-symmetry of the hybrid model to be a good guide to the IR physics, we need $V$ to be a vertical vector field, i.e. $\cL_{V} \pi^\ast(\omega) = 0$ for all forms $\omega \in \Omega^\bullet(B)$, and in particular the $\GUL\times\GUR$ symmetries fix $B$ point-wise.  We denote a model where this is the case a \emph{good hybrid}.  
As we show in Appendix~\ref{app:VKill} this implies
\begin{align}
\label{eq:KVform}
V = \textstyle{\sum}_{i=1}^n q_i \phi^i\pp{\phi^i} + \text{c.c.}~
\end{align}
for some real charges $q_i$.  The $q_i$ have to be compatible with the transition functions defining $X \to B$, and since $\cL_V W = W$, and $W$ is polynomial in every patch, $q_i \in \Q_{\ge 0}$.
In a LG theory, i.e $B$ a point, standard results show that if the potential condition is satisfied then without loss of generality $0 < q_i \le 1/2$~\cite{Kreuzer:1992bi,Klemm:1992bx}.  More generally, the potential condition requires that $W(y^I,\phi)$, thought of locally as a LG potential for the fiber fields $\phi$ depending on the ``parameters'' $y^I$, should be non-singular in a small neighborhood of any generic point in $B$.  Hence, the range of allowed $q_i$ is the same for a hybrid theory as it is for LG models.

\subsubsection*{The orbifold action}
Our main interest in the hybrid SCFTs is for applications to supersymmetric 
compactification of type II or heterotic string theories.  For left-right symmetric theories this requires the existence of $\GUL\times\GUR$ symmetries with integral $\bq$, $\bqb$ charges of all (NS,NS) sector states~\cite{Banks:1987cy}.  Our hybrid theory, if it flows as expected  to a $c=\cb =9$ SCFT in the IR will not satisfy this condition.  Fortunately, the solution is the same as it is for Gepner models~\cite{Gepner:1987vz} or LG orbifolds~\cite{Vafa:1989xc,Intriligator:1990ua}: we gauge the discrete symmetry $\Gamma$ generated by $\exp[2\pi i J_0]$, where $J_0$ denotes the conserved $\GUL$ charge; since all fields have $\bq-\bqb \in \Z$, the orbifold by $\Gamma$ is sufficient to obtain integral charges.  

In the line bundle case with $q_i = n_i/d_i$ we then see that $\Gamma \simeq \Z_N$, with $N$ the least common multiple of $(d_1,\ldots,d_n)$.  Since $\Gamma$ is embedded in a continuous non-anomalous symmetry we expect the resulting orbifold to be a well-defined quantum field theory, and the resulting orbifold SCFT will be suitable for a string compactification.  

In addition to the introduction of twisted sectors and the projection, the orbifold has one important consequence for the physics of hybrid models:  it allows us to consider more general ``orbi-bundles,'' where the fiber in $X\to B$ is of the form $\C^n/\Gamma$, and the transition functions are defined up to the orbifold action.   For instance, we will examine a theory with $B = \P^3$ and $X = \cO(-5/2)\oplus\cO(-3/2)$, where the orbifold $\Gamma =\Z_2$ reflects both of the fiber coordinates.\footnote{A GLSM embedding of this hybrid model is given in section 2.5 of~\cite{Addington:2013gpa}~.}

\subsection{The quantum theory and the hybrid limit}
Having defined the classical hybrid model's Lagrangian and discussed its symmetries, we now discuss the quantum theory.
To orient ourselves in the issues involved, let's recall the case of (2,2) LG models --- the simplest examples of hybrids.  These theories have a Lagrangian description at some renormalization scale $\mu$ as a free kinetic term for chiral multiplets, and a superpotential interaction with dimensionful couplings $m$.  The theory is weakly coupled when $\mu \gg m$, and we can use the Lagrangian and (approximately) free fields to describe the theory.  The low energy limit $\mu\to 0$ is then strongly coupled, and while $W$ is protected by SUSY non-renormalization theorems, the kinetic term receives a complicated but irrelevant set of corrections.  There is by now overwhelming evidence that these do flow to the expected SCFTs, in accordance with the original proposals~\cite{Martinec:1988zu,Vafa:1988uu}, and computations of RG-invariant quantities allow us to use the weakly coupled $\mu \gg m$ description to describe \emph{exactly} the SCFT's (c,c) chiral ring and more generally the $\bQb$-cohomology.  Furthermore, the results extend to LGOs suitable for string compactification.\footnote{These typically have non-trivial (a,c) rings encoded in the twisted sectors, and that ring structure is not easy to access directly via the LG orbifold description.}

There is a small IR subtlety in using the weakly coupled LG description:  the theory at $W=0$ is non-compact and has all the usual difficulties associated to non-compact bosons.  This is of course not very subtle since the theory is free; however, more to the point, in using the weakly coupled description we still keep track of the R-charges and weights that follow from the superpotential and do not consider states supported away from the $W=0$ locus.

A more general hybrid theory has a similar structure, except that now there are two sorts of couplings:  the superpotential couplings $m/\mu$, as well as the choice of K\"ahler class on the base $B$.  Although the latter coupling is typically encoded in the kinetic D-term, it can also be expressed as a deformation of the twisted chiral superpotential.  Hence, the K\"ahler class and superpotential couplings do not receive quantum corrections.  Of course we do expect corrections to the D-terms, but these should be irrelevant just as they are in the LG case.  Moreover, there is good evidence, based on GLSM constructions, that the hybrid models with a GLSM UV completion should flow to SCFTs with expected properties (i.e. correct central charges and R-symmetries), and we expect the same to hold for more general hybrid models.  As in the LG case, the strict $W=0$ limit may be subtle, perhaps even more so, since it may require us to specify additional details about the geometry of $\bY$.  However, we may use the same cure for these IR subtleties as we do in the LG case:  use the R-charges and weights encoded by the superpotential and restrict attention to field configurations and states supported on $B$.

Assuming a hybrid model does flow to an expected SCFT, we would like to have techniques to evaluate RG-invariant quantities such as the $\bQb$-cohomology.  It is here that there will be important conceptual and technical differences from the LG case due to the non-trivial base geometry $B$.  For instance, we expect the $\bQb$-cohomology to depend on the choice of K\"ahler class on $B$.  While there will not be a perturbative dependence, we do in general expect corrections from world-sheet instantons wrapping non-trivial cycles in $B$.  These corrections are suppressed when $B$ is large, which leads us to define the hybrid analogue of the large radius limit of a NLSM:  \emph{the hybrid limit}, where the K\"ahler class of $B$ is taken to be arbitrarily deep in its K\"ahler cone.  In what follows, we will study the $\bQb$-cohomology of a hybrid model in the hybrid limit.  

\section{Massless spectrum of heterotic hybrids} \label{s:spectra}
In this section we develop techniques to evaluate the massless spectrum for 
a compactification of the $\GE_8\times\GE_8$ heterotic string based on a $c=\cb=9$ (2,2) hybrid SCFT.\footnote{The $\SO(32)$ case can be handled in an entirely analogous fashion.}  We first review the standard prescription~\cite{Gepner:1987vz,Vafa:1989xc,Kachru:1993pg} to obtain a modular invariant theory and identify world-sheet Ramond ground states with massless fermions in spacetime.  We then discuss how to enumerate these ground states by studying the $\bQb$ cohomology in the hybrid limit.

\subsection{Spacetime generalities}
In order to describe a heterotic string compactification, we complete our hybrid $c=\cb=9$ $N=(2,2)$ SCFT internal theory to a critical heterotic theory by adding ten left-moving fermions (with fermion number $F_\lambda$) that realize an $\so(10)$ level $1$ current algebra, a left-moving level $1$ hidden $\Le_8$ current algebra, and the free $c=4$, $\cb = 6$ theory of the uncompactified spacetime $\R^{1,3}$.

A modular invariant theory is obtained by performing left- and right- GSO projections.
The left-moving GSO projection onto $e^{i\pi J_0} (-)^{F_\lambda} = 1$ is responsible for enhancing the linearly realized $\Lu(1)_{\text{L}}\oplus\so(10)$ gauge symmetry to the full $\Le_6$.  The right-moving GSO projection has a similar action, combining $\Jb_0$ with the fermion number of the $\R^{1,3}$ theory.  Its immediate spacetime consequence is $N=1$ spacetime supersymmetry, or equivalently, a relation, via spectral flow, between states in right-moving Neveu-Schwarz and Ramond sectors.  Spacetime fermions arise in the  (NS,R) and (R,R) sectors, and supersymmetry allows us to identify the full spectrum of supermultiplets in the spacetime theory from these states.  

The spacetime theory obtained by this procedure will have a model-independent set of massless fermions:  the gauginos of the hidden $\Le_8$, the gravitino, and the dilatino.  In what follows we focus on the model-dependent massless spectrum.  In particular, the hidden $\Le_8$ degrees of freedom are always restricted to their NS ground state and just make a contribution to the left-moving zero-point energy.

On-shell string states have vanishing left- and right-moving energies.   For massless states there is no contribution to $\Lb_0$ from the $\R^{1,3}$ free fields; massless fermions are thus states in the (R,R) and  (NS,R) sectors with vanishing left-moving and right-moving energies.  In the  (R,R) sector, massless states are associated to the ground states in the internal theory, related by spectral flow to (NS,NS) operators comprising the ``chiral rings'' \cite{Vafa:1988uu} of the theory.  Massless states in the (NS,R) sector include states related to these by left-moving spectral flow as well as additional states.  The main result of \cite{Kachru:1993pg} is a method for describing these states in LGO theories, which we here extend to hybrids.  This relies on the familiar fact that since 
\begin{align}
\{\bQ,\bQb\} = 2\Lb_0~;\quad \bQ^2 = \bQb^2 = 0
\end{align}
the kernel of $\Lb_0$ is isomorphic to the cohomology of $\bQb$.

The right-moving GSO projection is onto states with $\bqb\in \Z+\ff{1}{2}$; those with $\bqb=-1/2$ ($\bqb=1/2$) correspond to chiral (anti-chiral) multiplets, while states with $\bqb=\pm 3/2$ are gauginos in vector multiplets.  The $\GUL$ charge $\bq$ determines the $\Le_6$ representation according to the decomposition
\begin{align}\label{eq:E6decomposition}
&\Le_6 \supset \so(10)\oplus \Lu(1)\nonumber\\
& \mathbf{78} = \mathbf{45}_0 \oplus \mathbf{16}_{-3/2} \oplus  \mathbf{\overline{16}}_{3/2}  \oplus \mathbf{1}_0    \nonumber\\
& \mathbf{27} =  \mathbf{16}_{1/2} \oplus  \mathbf{{10}}_{-1}  \oplus \mathbf{1}_2    \nonumber\\
& \mathbf{\overline{27}} =  \mathbf{\overline{16}}_{-1/2} \oplus  \mathbf{{10}}_1  \oplus \mathbf{1}_{-2}~.
\end{align}

As in the LG orbifold case~\cite{Vafa:1989xc,Kachru:1993pg}, the GSO projection can be combined with the hybrid orbifold of $\Gamma=\Z_N$ to an orbifold by $\Z_2\ltimes\Z_N\cong \Z_{2N}$. Therefore we need to study the $2N$ sectors twisted by $[\exp(i\pi J_0)]^k, k=0,\dots,2N-1$.\footnote{That is, schematically, in the $k$-th twisted sector fields satisfy $\phi(ze^{2\pi i},\zb e^{-2\pi i}) = [\exp(i\pi J_0)]^k\phi(z,\zb)$.  We will make these periodicities more precise shortly.} Spacetime CPT exchanges the $k$-th and the $(2N-k)$-th sectors, and CPT invariance means we can restrict our analysis to the $k=0,1,\dots,N$. sectors.  The states arising in (R,R) ($k$ even) sectors give rise to $\Le_6$-charged matter.  This is easy to see since in this case the ground states of the $\so(10)$ current algebra transform in $\rep{16}\oplus\brep{16}$.  Massless $\Le_6$-singlets are of particular interest, and they can only arise from (NS,R) sectors, i.e. sectors with odd $k$.

\subsection{Left-moving symmetries in cohomology} \label{ss:n2alg}
The action of $\GUL$ commutes with $\bQb$, and following~\cite{Witten:1993jg,Silverstein:1994ih}, we can find a representative for the corresponding conserved current in $\bQb$-cohomology, denoted by $H_{\bQb}$.  Consider the operator
\begin{align}
\cJ_L \equiv \cX^\beta(D_\beta V^\alpha -\delta^\alpha_\beta)\cXb_{\alpha}- V^\alpha g_{\alpha\betab} \pz\Yb^{\betab}~.
\end{align}
Using~(\ref{eq:02eom}) and $\cL_{V} W = W$ it follows $\cDb \cJ_L = 0$.  Observing that $\cQb$ and $\cDb$ are conjugate operators, $\cQb = -\exp\left[2\thetab\theta \pbz\right] \cDb \exp\left[2\theta\thetab \pbz\right]$,
we conclude that 
\begin{align}
J_L \equiv \left.\cJ_L\right|_{\theta=0} = \chi^\beta(V^\alpha_{~,\beta} -\delta^\alpha_\beta)\chib_{\alpha}- V^\alpha\rho_\alpha
\end{align}
is $\bQb$-closed and hence has a well-defined action on $H_{\bQb}$.  Similarly, we can obtain the remaining generators of the left-moving $N=2$ algebra in $H_{\bQb}$.
To find the energy-momentum generator $T$ we observe that
\begin{align}
\cT_0 = -g_{\alpha\betab} \pz Y^\alpha \pz \Yb^{\betab} -\cX^\alpha \Dz \cXb_{\alpha} = -\pz Y^\alpha \left[g_{\alpha\betab}\pz\Yb^{\betab} -  g_{\gamma\betab,\alpha} \cX^\gamma \cXb^{\betab}\right] - \cX^\alpha\pz\cXb_\alpha
\end{align}
satisfies $\cDb \cT_0 = 0$, as does 
\begin{align}
\cT \equiv \cT_0 - \half\pz\cJ_L~.
\end{align}
The lowest component of $\cT$ is $\bQb$-closed and given by
\begin{align}
T &= -\p y^\alpha\rho_\alpha- \half\left(\chib_\alpha  \pz \chi^\alpha+ \chi^\alpha \pz\chib_\alpha\right) -\half\pz\left[ \chi^\beta \chib_\alpha V^\alpha_{~,\beta} -V^\alpha \rho_\alpha\right]\ .
\end{align}
The remaining generators of a left-moving $N=2$ algebra are obtained from the $\cDb$-closed fields
\begin{align}
\cG^+  \equiv i\sqrt{2} \left[ \cXb_{\alpha} \pz Y^\alpha - \pz(\cXb_{\alpha} V^\alpha)\right],\qquad
\cG^- \equiv i\sqrt{2} \left[ \cX^\alpha g_{\alpha\betab} \pz\Yb^{\betab}\right]~,
\end{align}
yielding the left-moving supercharges $G^{\pm}$ in $H_{\bQb}$:
\begin{align}
G^+ = i\sqrt{2} \left[ \chib_{\alpha} \pz y^\alpha - \pz(\chib_{\alpha} V^\alpha)\right]~,\qquad
G^- \equiv i\sqrt{2} \chi^\alpha \rho_\alpha~.
\end{align}

\subsection{Reduction to a curved $bc-\beta\gamma$ system} \label{ss:bcbg}
The action~(\ref{eq:rhoaction}) determines the OPEs for the left-moving degrees of freedom to be
\begin{align}
\label{eq:OPEs}
y^\alpha(z) \rho_{\beta}(w) \sim \frac{1}{z-w}\delta^\alpha_\beta~,\qquad
\chi^\alpha(z)\chib_{\beta}(w) \sim \frac{1}{z-w}\delta^\alpha_\beta~.
\end{align}
Using the normal ordering defined by these free-field OPEs we can define $T$, $J$, and $G^\pm$ in the quantum theory.  This is particularly simple with our choice of fields and Killing vector $V$: the operators are quadratic in the fields, and it is easy to check that they indeed generate an $N=2$ algebra with central charge
\begin{align}
\label{eq:centralcharge}
c = 3 d + 3\textstyle{\sum}_{i=1}^n (1-2q_i)~,
\end{align}
which we recognize as the sum of the fiber LG central charge and the contribution from the base.  The $\GUL$ charge $J_0$ and left-moving Hamiltonian $L_0$ are obtained in the standard fashion as
\begin{align}
\label{eq:leftcharges}
J_0 = \oint \frac{dz}{2\pi i} J_L(z)~,\qquad L_0 = \oint \frac{dz}{2\pi i} z T(z)~,
\end{align}
and the resulting charge and weight assignments for the fiber fields are given in table~\ref{table:charges} together with the $\GUR$ charge $\bqb$.  
\begin{table}[t]
\begin{center}
\label{table:charges}
\begin{tabular}{|c|c|c|c|c|c|c|c|c|}
\hline
~ 	
&$y^I$			&$\rho_I$				&$\chi^I$			&$\chib_I$	
&$\phi^i$			&$\rho_i$				&$\chi^i$			&$\chib_i$  \\ \hline
$\bq$	
&$0$		&$0$		&$-1$	&$1$
&$q_i$		&$-q_i$			&$q_i-1$			&$1-q_i$	\\ \hline
$2h$
&$0$		&$2$		&$1$		&$1$	
&$q_i$	&$2-q_i$	&$1+q_i$ 	& $1-q_i$ \\ \hline
$\bqb$ 
&$0$			&$0$				&$0$				&$0$
&$q_i$		&$-q_i$			&$q_i$			&$-q_i$ 
\\ \hline
\end{tabular}
\caption{Weights and charges of the fields.}
\end{center}
\end{table}
These currents are trivially annihilated by $\bQb_0$ and commute with $\bQb_W$, whose action is now realized as
\begin{align}
\label{eq:bQbWop}
\bQb_W \equiv \oint \frac{dz}{2\pi i} ~ \left[ \chi^\alpha W_\alpha (y)\right] (z)~.
\end{align}

It may seem a little bit puzzling that we have been able to reduce the entire problem to a free first order system.  What, the reader may ask, encodes the target space geometry, for example?  The answer, familiar from~\cite{Nekrasov:2005wg,Witten:2005px}, is that the free field theory description only applies patch by patch in field-space.  That is, we cover $\bY$ with open sets $U_{a}$ and local coordinates $x_a^\alpha$, and on each $U_{ab} = U_{a}\cap U_{b} \neq \emptyset$  $x_b=x_b(x_a)$, and we define the holomorphic transition functions
\begin{align}
(T_{ba})^\alpha_{\beta} \equiv \frac{\p x^\alpha_b}{\p x^\beta_a},\qquad
(\cS_{ba})^\alpha_{\beta\gamma} \equiv (T_{ba}^{-1})^\alpha_\delta (T_{ba})^\delta_{\beta,\gamma}~.
\end{align}
The left-moving fields then patch according to
\begin{align}
\label{eq:patchfields}
y^\alpha_b &= x^\alpha_b(y_a)~,\quad
\chi^\alpha_b = (T_{ba})^{\alpha}_{\beta}\chi^\beta_a~,\quad
\chib_{b\alpha} = (T_{ba}^{-1})^\beta_\alpha \chib_{a\beta}~,\nonumber\\
\rho_{b\alpha} &= : (T_{ba}^{-1})^\beta_{\alpha}\left[ \rho_{b\beta} -\cS^\delta_{ba\beta\gamma} \chib_\delta\chi^\gamma\right]:~,
\end{align}
where the transition functions are evaluated at $y_a$, e.g. $T_{ba} = T_{ba}(y_a)$.  Note that the patching of $\rho$ requires a normal-ordering due to singularities in the $y-\rho$ and $\chib-\chi$ OPEs.  Of course there are similar transformations for the right-moving fields $\yb$ and $\eta,\etab$.  For instance, the $\etab^{\Ib}$ transform as sections of $y^\ast(\Tb_B)$.\footnote{As we are working on a flat world-sheet throughout this paper, we do not keep track of the world-sheet spinor properties of the fermionic degrees of freedom.}

These transition functions require a careful analysis when we expand about world-sheet instanton configurations, i.e. non-trivial holomorphic maps $\Sigma \to \bY$.  This, together with non-trivial fermi zero modes in the background of an instanton will lead to world-sheet instanton corrections to $\bQb_0$.\footnote{Since $\bQb_W$ is associated to a chiral superpotential, we do not expect it to be corrected by world-sheet instantons.}  These corrections vanish in the hybrid limit where we expand about constant maps $\pz y = \pbz y = 0$, and the only non-trivial $\bQb_0$ action is on the anti-holomorphic zero modes $\bQb_0 \cdot \yb^{\alphab}_0 = -\etab^{\alphab}_0$.   In fact, since the $\etab^{\ib}$ are $\bQb$-exact, as far as cohomology is concerned, we can safely ignore the $\etab^{\ib}$ as well as the anti-holomorphic bosonic fiber zero modes $\phib^{\ib}_0$.  So, the only non-trivial $\bQb_0$ action is on the base anti-holomorphic zero modes: $\bQb_0 \cdot \yb^{\Ib}_0 = -\etab^{\Ib}_0$.  In what follows we will drop the zero mode subscript on these right-moving fields with the understanding that $\yb$ and $\etab$ will  denote the base antiholomorphic zero modes.

\subsection{Massless states in the hybrid limit} \label{ss:massless}
Our task now is to work out, in each twisted sector, the set of GSO-even states that belong to $H_{\bQb}$ and carry left-moving energy $E= 0$. We construct the relevant states (i.e. the only ones with required energy and charges) in the Hilbert space as polynomials in the fermions and non-zero bosonic oscillator modes tensored with wavefunctions of the bosonic zero modes.  In a generic twisted sector the bosonic zero modes correspond to the compact base $B$, while in less generic sectors there can be additional bosonic zero modes.  However, since the non-compact bosonic modes will be lifted by the superpotential, in what follows all bosonic wavefunctions will be taken to be polynomial in the fiber fields.  

   The operators $T$ and $J_L$ can be used to grade the states according to their energy $E$ and left-moving charge $\bq$, and we can evaluate $\bQb$-cohomology on the states of fixed $E$ and $\bq$.
   An important simplification comes from working in the right-moving Ramond ground sector.  A look at~(\ref{eq:supersplit}) shows that, as far as $\bQb$-cohomology is concerned, we can neglect any states containing oscillators in $\pbz y^\alpha$, as well as any non-zero mode of $\eta^\alpha$.  We choose the Ramond ground state annihilated by the zero modes of $\eta^\alpha$, so our states will be constructed without $\eta^\alpha$ or right-moving bosonic oscillators.   We will call the resulting space of states  \textit{ the restricted Hilbert space} $\cH$.  In general this will be infinite-dimensional even at fixed $E$ and $\bq$.

\subsubsection*{Twisted modes and ground state quantum numbers}
In this section we provide expressions for $E$, $\bq$ and $\bqb$ of the states in a fixed twisted sector.   For simplicity, we work out the case $X = \oplus_i L_i$.  The result extends immediately to orbi-bundles of the form $X = \oplus_i L_i^{x_i}$ for $x_i \in \Q$.  It should be possible to treat the case of more general $X$ at the price of additional notation.

The first task is to describe the mode expansions of the fields and the quantum numbers of the ground states $|k\ra$.
While we can restrict to right-moving (i.e. anti-holomorphic) zero modes, the left-moving oscillators need to be treated in detail.  In each patch of the target space the moding of the left-moving fields in the $k$-th twisted sector is
\begin{align}
\label{eq:modes}
y^\alpha(z)&=\sum_{r\in\Z-\nu_\alpha} y^\alpha_r z^{-r-h_\alpha} , \qquad \qquad \quad   \chi^\alpha(z)=\sum_{r\in\Z-\nut_\alpha} \chi^\alpha_r z^{-r-\htld_\alpha}, \nonumber\\
\rho_\alpha(z)&=\sum_{r\in\Z+\nu_\alpha} {\rho_\alpha}_r z^{-r+h_\alpha-1} , \qquad \qquad   \chib_\alpha(z)=\sum_{r\in\Z+\nut_\alpha} \chib_{\alpha r} z^{-r+\htld_\alpha-1},
\end{align}
where
\begin{align}\label{eq:nus}
\nu_\alpha&=\frac{kq_\alpha}{2} \mod1~, &\nut_\alpha&=\frac{k(q_\alpha-1)}{2} \mod1~,&
\htld_{\alpha} -\frac{1}{2} &= h_\alpha = \frac{q_\alpha}{2}~.
\end{align}
We choose $0\leq \nu_\alpha<1$ and $-1<\nut_\alpha\leq 0$ and recall
that the oscillator vacuum $|k\ra$ is annihilated by all the positive modes.  When $\chi,\chib$ have zero modes our conventions are that the ground state is annihilated by the $\chi_0$ modes.  

The mode (anti)commutators follow from~(\ref{eq:OPEs}) and~(\ref{eq:modes}):
\begin{align}
\CO{y^\alpha_r}{\rho_{\beta s}} = \delta^{\alpha}_{\beta} \delta_{r,-s}~,\qquad
\AC{\chi^\alpha_r}{\chib_{\beta s}} = \delta^{\alpha}_{\beta} \delta_{r,-s}~.
\end{align}
Each oscillator carries the obvious $\bq,\bqb$ charges and contributes minus its mode number to the energy.  By using this mode expansion to compute $1$-point functions  of $T$ and $J_L$ in the oscillator vacuum $|k\ra$, we determine the quantum numbers of $|k\ra$.  The left- and right-moving charges are given by
\begin{align}
\bq_{|k\ra}&=\sum_\alpha \left[ (q_\alpha-1)(\nut_\alpha+\half)-q_\alpha(\nu_\alpha-\half)   \right] , \nonumber\\
\bqb_{|k\ra}&=\sum_\alpha \left[ q_\alpha(\nut_\alpha+\half)+(q_\alpha-1)(-\nu_\alpha+\half)   \right]  ,
\end{align}
and while the left-moving energy  is $E_{|k\ra} =0$ for $k$ even, we have 
\begin{align}
E_{|k\ra} = -\frac{5}{8} + \half \sum_\alpha \left[ \nu_\alpha(1-\nu_\alpha)+\nut_\alpha(1+\nut_\alpha)   \right] ,
\end{align}
for $k$ odd.  Note that this includes the usual $-c/24$ shift: $E=L_0-1$.  

The oscillator vacuum $|k\ra$ we have constructed is not in general a state in the Hilbert space.   To specify a state we need to prescribe a dependence on the bosonic zero modes so as to get a well-defined state, but from above we see that $|k\ra$ 
transforms as a section of a holomorphic line bundle $L_{|k\ra} $ over $B$.   When $X=\oplus_iL_i$ we find (using $K_{\bY} = \cO_{\bY}$)
\begin{align}\label{eq:bundlegroundst}
L_{|k\ra}=\begin{cases}
 \otimes_i L_i^{(\nut_i-\nu_i)}   &\qquad  \text{for } k \text{ even}, \\
 \otimes_i L_i^{(\nut_i-\nu_i+\half)} & \qquad \text{for } k \text{ odd}~.
\end{cases}
\end{align}
From (\ref{eq:nus}) we see that if we set $\nu_I=0$ and $\nut_I = -k/2\mod 1$, then $\tau_\alpha = \nu_\alpha - \nut_\alpha$ is 
\begin{align}
\label{eq:tau}
\tau_\alpha = \begin{cases} 
0     &\nu_\alpha = 0\\
1     &\nu_\alpha \ne 0
\end{cases}\quad\text{for even $k$}~;\quad
\tau_\alpha = \begin{cases} 
1/2 & \nu_\alpha\le \half\\
3/2 & \nu_\alpha > \half
\end{cases}\quad\text{for odd $k$}~.
\end{align}
This shows that $L_{|k\ra}$ is well-defined because $\tau_\alpha\in \Z$ for $k$ even and $\tau_\alpha\in \Z+\half$ for $k$ odd.   A well-defined ground state can be of the form 
\begin{align}
|\Psi^k_0\ra = \Psi_0(y',\yb)_{\Ib_1\cdots \Ib_u}\etab^{\Ib_1}\cdots\etab^{\Ib_u}|k\ra\ ,
\end{align}
where $y'$ denotes bosonic zero modes, the $\etab^{\Ib}$ are the right-moving superpartners of the base coordinates and $\Psi^k_u$ are (0,u) horizontal forms on $\bY$ valued in the holomorphic sheaf $L_{|k\ra}^\ast$.
In sectors in which there are additional zero modes ($k=0$ is always an example of this) there are more general ground states, and in (R,R) sectors a subset of these ground states describes the massless spectrum. 

This non-trivial vacuum structure is a generalization of familiar limiting cases of the hybrid construction.  When $\bY = B$ a compact Calabi-Yau manifold, the Ramond ground state is a section of a trivial bundle (the square root of the trivial canonical bundle); in  the LGO case each twisted sector has a unique ground state $|k\ra$.

\subsubsection*{The double-grading and spectral sequence}
Our restricted Hilbert space $\cH$ at fixed $E$ and $\bq$ admits a grading by $\GUR$ charge, and $\bQb$ acts as a differential, $\bQb: \cH_{\bqb} \to \cH_{\bqb+1}$ that preserves the left-moving quantum numbers.  A key observation, made in the LG case in~\cite{Kachru:1993pg}, that makes the cohomology problem tractable is that in fact $\cH$ admits a double-grading compatible with the split $\bQb = \bQb_0+ \bQb_W$ in~(\ref{eq:supersplit}).  Let $U$ be an operator that assigns charge $+1$ to $\etab$, $-1$ to $\eta$, and leaves the other fields invariant. Although  $U$ is not a symmetry of the theory when $W\neq 0$, we can still grade our restricted Hilbert space according to the eigenvalues $u$ of $U$ and $p\equiv \bq-u$, and since $\CO{U}{\bQb_0} = \bQb_0$ and $\CO{U}{\bQb_W} = 0$ we obtain a double-graded complex with
\begin{align}
\bQb_0 : \cH^{p,u} \to \cH^{p,u+1},\qquad
\bQb_W :\cH^{p,u} \to \cH^{p+1,u}~
\end{align}
acting, respectively, as anticommuting vertical and horizontal differentials.  The cohomology of $\bQb$ is thus computed by a spectral sequence with first two stages
\begin{align}
E_1^{p,u}=H^u_{\bQb_0}(\cH^{p,\bullet}),\qquad\text{and}\qquad E_2^{p,u}=H^p_{\bQb_W}H^u_{\bQb_0}(\cH^{\bullet,\bullet})~.
\end{align}
In general, $E_{r+1}$ is obtained from $E_r$ as the cohomology of a differential $d_r$ acting as
\begin{align}
d_r:E_r^{p,u}\to E_r^{p+r,u+1-r}\ .
\end{align}
We have, for example, $d_0=\bQb_0$ and $d_1 = \bQb_W$.  The differentials at higher stages are produced by a standard zig-zag construction~\cite{Bott:1982df}.
Since the range of $U$ is $0\le U\le d$ the differentials vanish for $r>\dim B$, and the sequence converges:
$E_{\dim B+1}^{p,u}=E_\infty^{p,u}=H^{p,u}_{\bQb}(\cH^{\bullet,\bullet})$.

We now have almost all of the tools to describe the massless spectrum.  In each twisted sector there is a geometric structure that organizes the states in the spectral sequence.  On $\cH$ the $\bQb_0$ action is simply
\begin{align}
\bQb_0 = -\etab^{\Ib} \frac{\p}{\p \yb^{\Ib}},
\end{align} 
so $\bQb_0$ cohomology amounts to restricting to horizontal\footnote{We mean in the sense of the fiber--base geometry of $\bY$.} Dolbeault cohomology groups, while $\bQb_W$ cohomology imposes further algebraic restrictions.  

Since the geometry is typically non-compact the $\bQb_0$ cohomology groups are often infinite-dimensional.  Fortunately we can obtain a well-defined counting problem because $\bQb_0$ respects the \emph{fine grading} by a vector $\br = (r_1,\ldots, r_n)\in \Z^n$ that assigns grade $\br$ to a monomial $\prod_i \phi_i^{r_i}$.\footnote{This grading has a simple physical interpretation:  the $W=0$  theory has $n$ $\GU(1)$ symmetries that rotate the fiber fields separately.}  Restricting to a particular grade leads to finite-dimensional vector spaces that, as we show in appendix~\ref{app:sheaf}, are readily computable in terms of sheaf cohomology over $B$.
The fine grading is a refinement of the physically relevant grading by $\bq$ and $E$, and therefore it gives an effective method for evaluating the first stage in the spectral sequence $E_1^{p,u}$ at fixed twisted sector, $\bq$, and $E$.

The next step is to study the $\bQb_W$ cohomology, i.e. the second stage $E_2^{p,u}=H^p_{\bQb_W}\left(H^u_{\bQb_0}(\cH^{\bullet,\bullet})\right)$.   Once the first two stages of the spectral sequence are determined, we are able to compute the cohomology of $\bQb$; higher derivatives are then determined by standard zig-zag arguments in terms of the two differentials $\bQb_0$ and $\bQb_W$.

The geometric structure depends on the twisted sector, and rather than presenting a universal framework at the price of opaque notation, we will next consider the relevant geometries in three separate situations:
\begin{enumerate}
\item The (R,R) sectors: $k \in 2\Z$.  In this case since $E_{|k\ra} = 0$ we can restrict to zero modes for all the fields, which leads to a very transparent structure.
\item The untwisted (NS,R) sector: $k=1$.  This and its CPT conjugate sector $k=2N-1$ are the only states with $E_{|k\ra} = -1$.  In this case the geometry is simply $\bY$, and the spectrum involves an interplay between non-trivial base and fiber oscillators.
\item (NS,R) sectors with odd $k$ and $E_{|k\ra} >-1$.  In this case the organizing geometry is a sub-bundle of $\bY \to B$, and while the choice of sub-bundle is $k$-dependent, the spectrum simplifies since base oscillators have $h=1$ and do not contribute to the massless states.
\end{enumerate}
We consider these possibilities in turn in the next section.

\section{Twisted sector geometry} \label{s:twisted}
To describe the geometric framework for the various twisted sectors  we find it useful to distinguish base and fiber fields, 
with the latter differentiated according to the values of $\tau_\alpha$.  More precisely, we split the coordinates $y^\alpha \to (y^{\alpha'},\phi^A)$, such that $\tau_{\alpha'} <1$ and $\tau_{A} \ge 1$.  The $y^{\alpha'}$ decompose further into base and fiber directions:  $y^{\alpha'} = (y^I,\phi^{i'})$, where $\tau_{i'} < 1$ (since $\nu_I =0$ for all the base fields $\tau_I< 1$ in all sectors).
We decompose the bundle $X$ accordingly as $X = X_k \oplus \oplus_A L_A$ and define 
\begin{align}
\label{eq:bYk}
\bYk \equiv \text{tot} ( X_k \overset{\pi_k}{\longrightarrow} B).
\end{align}  
The utility of this is that the ``light'' fields, labeled by $\alpha'$, including the corresponding fermions, are organized by $\bYk$, while the remaining ``heavy'' fields, labeled by $A$, are organized by the pull-backs $\pi^\ast_k(L_A)$.  The right-moving sector is considerably simpler: we restrict to zero modes, and as we described at the end of section~\ref{ss:bcbg}, the only relevant ones are the zero modes $\yb^{\Ib}$ and their $\bQb_0$ superpartners $\etab^{\Ib}$.  We now describe how this works in detail in various twisted sectors.

\subsection{(R,R) sectors}
In this case $E_{|k\ra} = 0$ as a consequence of the left-moving supersymmetry, and to describe the massless states we can restrict to zero modes for all the fields.  A look back at the modes in~(\ref{eq:modes}) and (\ref{eq:nus}) shows that the only fields with zero modes are the light fields.  Among these the $\rho_{\alpha'}$ also have no zero modes, while the $\chi_{\alpha'}$ zero modes annihilate the vacuum state.  Hence the most general state in the truncated Hilbert space is a linear combination of
\begin{align}
|\Psi^s_u \ra = \Psi(y',\xb)^{\alpha'_1 \cdots \alpha'_s}_{\Ib_1 \cdots \Ib_{u}} \chib_{\alpha'_1} \chib_{\alpha'_2}\cdots \chib_{\alpha'_s} \etab^{\Ib_1} \cdots \etab^{\Ib_u} |k\ra~.
\end{align} 
The fermions $\chib_{\alpha'}$ and $\etab^{\Ib}$ transform respectively as sections of $T^\ast_{\bYk}$ and $\pi^\ast_k(\Tb_{B})$,\footnote{The pull-back to the world-sheet is irrelevant since in the hybrid limit we consider constant maps.} while $|k\ra$ is a section of $L_{|k\ra} = \pi^\ast_k(\otimes_A L_A^\ast)$.  Hence to be a well-defined state the wavefunction $\Psi^s_u$ must be a $(0,u)$ horizontal form valued in the holomorphic bundle $\cE^s = \wedge^s T_{\bYk} \otimes L_{|k\ra}^\ast$.  

We can decompose the $\Psi$ according to their eigenvalues under the Lie derivative with respect to the restriction of the holomorphic Killing vector $V$ to $\bYk$ :  $\cL_{V} \Psi = q_\Psi \Psi$.\footnote{The Lie derivative has a well-defined action even when $L_{|k\ra}$ is non-trivial because $V$ is a vertical vector, while the transition functions for $L_{|k\ra}$ only depend on $B$.}  The resulting $|\Psi\ra$ has well-defined $\GUL\times\GUR$ charges:
\begin{align}
\bq = \bq_{|k\ra} + q_{\Psi} + s~,\qquad
\bqb = \bqb_{|k\ra} + q_{\Psi} + u~.
\end{align}
$\bQb_0$ acts by sending $\Psi^s_u \to -\pb \Psi^s_{u+1}$, and we can use the fine grading described in appendix~\ref{app:sheaf} to reduce $\bQb_0$ cohomology to computing the finite-dimensional vector spaces $H^\bullet_{\br}(\bYk,\cE^\bullet)$.

The result is still infinite-dimensional, since these cohomology groups will be non-zero for an infinite set of grades $\br$.  This is a general feature of any sector with bosonic fiber zero modes.  Fortunately, the action of $\bQb_W$, which takes the form
\begin{align}
\bQb_W = W_{\alpha'} (y') \chi^{\alpha'}~,
\end{align}
restricts the spectrum further.  When $W$ is non-singular we expect a finite-dimensional result, and indeed, this is easy to prove for LG models.\footnote{The result follows from the finite-dimensionality of the Koszul cohomology groups associated to the ideal $\la W_1,\cdots,W_n\ra \in \C[\phi_1,\ldots,\phi_n]$ for a non-singular superpotential~\cite{Kawai:1994qy,Melnikov:2009nh}.}  It would be useful to give a more general proof for hybrids.  At any rate, we see from~(\ref{eq:bQbWop}) that the $\bQb_W$ action on our state is simply
\begin{align}
\bQb_W : \Psi^s_u \mapsto (s W_{\alpha'_1} \Psi^{\alpha'_1 \alpha'_2 \cdots \alpha'_s})^{s-1}_u~.
\end{align}
The spacetime interpretation of these states is either as $\Le_6$ gauginos ($\bq = \pm 3/2$) or the $\rep{16}_{\pm 1/2}$ components of $\rep{27}$s and $\brep{27}$s.

\subsubsection*{$\bY = B$}
As a simple consistency check we can see that we correctly reproduce the expected spectrum from the unique $k=0$ (R,R) sector when $\bY = B$ a compact Calabi-Yau 3-fold.  The non-vanshing $\bQb_0$-cohomology classes, given with multiplicities and $(\bq,\bqb)$ charges are
\begin{align}
|0\ra_{-3/2,-3/2}^{\oplus 1} &&
|\Psi^3_0\ra_{3/2,-3/2}^{\oplus 1}&&
|\Psi^0_3\ra_{-3/2,3/2}^{\oplus 1}&&
|\Psi^3_3\ra_{3/2,3/2}^{\oplus 1}~, \nonumber\\[1.5mm]
|\Psi^1_1\ra_{-1/2,-1/2}^{\oplus h^1(T)}&&
|\Psi^2_2\ra_{1/2,1/2}^{\oplus h^1(T)}&&
|\Psi^2_1\ra_{1/2,-1/2}^{\oplus h^1(T^\ast)}&&
|\Psi^1_2\ra_{-1/2,1/2}^{\oplus h^1(T^\ast)}~.
\end{align}
Comparing to~(\ref{eq:E6decomposition}), we see that the first line corresponds to the gauginos, while the second line corresponds to the $\brep{16}_{-1/2}$ and $\rep{16}_{1/2}$ components of $h^1(T)$ chiral $\brep{27}$ and $h^1(T^\ast)$ chiral $\rep{27}$ multiplets.

\subsection{The $k=1$ sector} \label{ss:k1}
The $k=1$ sector is untwisted with respect to the LG orbifold action.  It  has the richest geometric structure and a number of universal features generalizing those observed for the LGO case~\cite{Aspinwall:2010ve}.  Since $\tau_\alpha = 1/2$ for all the fields, the geometry is simply $\bY_{\!\! 1} = \bY$, while the vacuum bundle $L_{|k\ra} = K_{\bY}$ is trivial.  We also have
\begin{align}
\bq_{|1\ra} = 0~,\qquad \bqb_{|1\ra} = -3/2~,\qquad E_{|1\ra} = -1~.
\end{align}
Since $E_{|1\ra} = -1$ massless states may include non-zero modes of $\p y^I$ and $\rho_I$.

We now want to describe the operators that create zero-energy states from $|1\ra$.  It turns out that hybrid theories for which some $q_i =1/2$ have additional zero-energy states that are not found in more generic theories.  We will first describe the zero energy states present generically and then turn to the special states available due to fields with $q_i =1/2$.

\subsubsection*{Generic $k=1$ operators} 
Ignoring multiplets with $q_i = 1/2$, we list the operators that can carry weight $h\le1$:\footnote{Working with fields, as opposed to modes, avoids complications in patching the non-trivial bosonic oscillators on the base.  These complications do not arise in sectors with $E_{|k\ra} >-1$.} 
\begin{align}
\label{eq:k1ops}
\cO^{1,s} &= \Psi^{1s\alpha_1\cdots\alpha_s}(y)\chib_{\alpha_1}\cdots\chib_{\alpha_s}~,\qquad
\cO^2 =\Psi^2_{\alpha}(y) \chi^\alpha~,\qquad 
\cO^3 = \Psi^3_{\alpha\beta}(y) \chi^\alpha\chi^\beta~,\qquad
\nonumber\\[1.5mm]
\cO^4 &= \Psi^4_{\alpha}(y) \p y^\alpha~,\qquad\qquad\qquad\quad~~
\cO^5 = ~:\Psi^{5\alpha}_{\beta}(y) \chib_\alpha \chi^\beta :~,\nonumber\\[1.5mm]
\cO^6 &=~ : \Psi^{6\alpha}(y) \rho_\alpha + \Psi^{6\alpha}_{~,\beta}(y) \chib_\alpha \chi^\beta :~.
\end{align}
The index $s$ in $\cO^{1s}$ can take values $s=0,1,2,3$.  In each case we only indicated the dependence on the left-moving fields; each $\Psi$ also depends on the $\yb$ and $\etab$ zero modes:
\begin{align}
\label{eq:wavefuexp}
\Psi^t = \sum_{u=0}^d (\Psi^t_{u})_{\Ib_1 \cdots \Ib_u} \etab^{\Ib_1} \cdots \etab^{\Ib_u}~,
\end{align}  
and plugging in this expansion, we obtain a set of operators $\cO^t_u(z)$.
We also used the normal ordering that follows from~(\ref{eq:OPEs}) to subtract off the $y\rho$ and $\chib\chi$ short-distance singularities.  Since our free fields are only defined on open sets covering the target space $\bY$, just as in the $k$ even case the wavefunctions $\Psi^t_0$ have to transform as sections of appropriate holomorphic bundles $\cE^t$ over $\bY$.  For instance, the fermi bilinear term appearing in $\cO^8$ is chosen to account for the
unusual transition function of $\rho_\alpha$ in~(\ref{eq:patchfields}).  That is, using~(\ref{eq:patchfields}), we find that for two patches $U_a$ and $U_b$ with $U_{ab} \neq \emptyset$ $\cO^6_{b} = \cO^6_{a}$ (i.e. $\cO^6$ is well-defined) iff $\Psi^6_0$ transforms as a section of $T_{\bY}$.  Similarly, the remaining wavefunctions must transform in the expected way, e.g. $\Psi^{1s}_0$ as  a section of $\wedge^s T_{\bY}$ and $\Psi^2_0$ as a section of $T^\ast_{\bY}$.  The wavefunctions for $\Psi^t_{u>0}$ transform as (0,u) horizontal forms valued in $\cE^t$, and taking $\bQb_0$ cohomology means the $\Psi^t_u$ taken at a fine grade $\br$ define classes in $H^\bullet_{\br}(\bY,\cE^\bullet)$.  As in the $k$ even case we need to consider all $\br$ that contain states with $h=1$ and non-trivial $\bQb_W$ classes.  It is useful to introduce the following notation for the relevant holomorphic bundles $\cE^t$:
\begin{align}
\label{eq:Bdef}
B_{s,t,q} \equiv \wedge^s T_{\bY} \otimes \wedge^t T^\ast_{\bY} \otimes \Sym^q (T_{\bY})~.
\end{align}

If we grade the wavefunctions by the eigenvalue of the Lie derivative with respect to the symmetry vector $V$, i.e. $\cL_{V} \Psi^t_u = q \Psi$, then we obtain the following weights, charges and $\bQb_W$ action for these operators: 
$\bqb_{\cO}= q+ u$, and 
\begin{align}
\label{eq:eq:k1ops2}
\xymatrix@R=1.5mm@C=1.5mm{
\text{op.} 	& \cO^{1,s}_u	&\cO^2_u	&\cO^3_u	&\cO^4_u	&\cO^5_u	&\cO^6_u\\
\bq_{\cO}		&q+s		&q-1		&q-2		&q		&q		&q		\\
h_{\cO}		&\frac{q+s}{2}
					&\frac{q+1}{2}
							&\frac{q+2}{2}
									&\frac{q+2}{2}
											&\frac{q+2}{2}
													&\frac{q+2}{2}\\
\bQb_W\cdot	
	&sW_{\alpha_1}\Psi^{1s\alpha_1\cdots\alpha_s}\chib_{\alpha_2}\cdots\chib_{\alpha_s}		
						&0		&0		&0
											&\Psi^{5\beta}_{u\gamma}W_\beta\chi^\gamma
													&\chi^\alpha\p_\alpha(\Psi^{6\beta}W_\beta)
}
\end{align}
Note that for $s>0$ the $\cO^{1,s}$ can carry negative eigenvalues under $\cL_V$, but it is not hard to show that they are bounded by $q >-s/2$.
Using these operators we create states in the usual fashion: $|\cO^{t}_u\ra  \equiv \lim_{z\to 0} \cO^{t}_u(z) |1\ra$.  They carry energy $E=h_{\cO}-1$ and charges $\bqb = \bqb_{\cO}-3/2$ and $\bq = \bq_{\cO}$.

\subsubsection*{Currents}
The $h_{\cO}=1$ $\bqb_{\cO}=0$ operators in $\bQb$ cohomology are conserved left-moving currents, and in a generic $k=1$ sector the corresponding states arise in the bottom row of the spectral sequence:
\begin{align}
\label{eq:currents}
\xymatrix{|\cO^5_0\ra \oplus |\cO^6_0\ra \ar[r]^-{\bQb_W} & |\cO^2_0\ra}~,  
\end{align}
where 
\begin{align}
\Psi^5 &\in \oplus_{\br} H^0_{\br} (\bY,B_{1,1,0})~,&
\Psi^6 &\in\oplus_{\br} H^0_{\br}(\bY,B_{0,0,1})~,& 
\Psi^2\in \oplus_{\br} H^0_{\br}(\bY,B_{1,0,0})~.
\end{align}
Before taking cohomology, there are a number of states here, including, for example, holomorphic vector fields in $H^0(B,T_B)$ that lift to $\bY$ or various enhanced R-symmetries of the $W=0$ theory.  Most of these states are lifted by the superpotential couplings.  In fact, for a suitably generic $W$ the only current that survives is $J_L$, which corresponds to $\Psi^{5} = \iden$ and $\Psi^{6} = - V$; the resulting state is $\bQb_W$ closed as a result of $\cL_V W = W$.  This gaugino corresponds to the linearly realized $\Lu(1)_L\subset \Le_6$.  For less generic $W$ additional currents may appear, and of course they are accompanied by additional chiral $\bqb=-1/2$ states $|\cO^2_0\ra$ in the cokernel of $\bQb_W$.  In spacetime each current corresponds to a gauge boson, and the appearance of extra currents reflects the spacetime Higgs mechanism.

\subsubsection*{$\bY = B$}
As in the $k=0$ case, we examine the case of trivial fiber and a CY target space.  Taking $\bQb_0$ cohomology on the space of operators in~(\ref{eq:k1ops}), we find the following massless states with $\bqb < 0$ (for brevity we omit their conjugates with $\bqb>0$)
\begin{align*}
\cO^{1,0},\cO^5_0 
& \to |1\ra_{0,-3/2}^{\oplus 1} \oplus \chib_\alpha \chi^\alpha|1\ra_{0,-3/2}^{\oplus 1}~
&&
\rep{45}_0\oplus\rep{1}_0 
\nonumber\\[1mm]
\cO^{1,1},\cO^2 
& \to |\cO^{1,1}_1\ra_{1,-1/2}^{\oplus h^1(T)}\oplus|\cO^2_1\ra_{-1,-1/2}^{\oplus h^1(T^\ast)}~
&&\rep{10}^{\oplus h^1(T)}_1\oplus\rep{10}_{-1}^{\oplus h^1(T^\ast)} 
\nonumber\\[1mm]
\cO^{1,2},\cO^3 
& \to |\cO^{1,2}_1\ra_{2,-1/2}^{\oplus h^1(\wedge^2T)} \oplus|\cO^3_1\ra_{-2,-1/2}^{\oplus h^1(\wedge^2T^\ast)}~
&&\rep{1}_{2}^{\oplus h^1(T^\ast)}\oplus\rep{1}_{-2}^{\oplus h^1(T)} \nonumber\\[1mm]
\cO^4,\cO^5_1,\cO^6 & \to |\cO^4_1\ra_{0,-1/2}^{\oplus h^1(T^\ast)} \oplus|\cO^5_1\ra_{0,-1/2}^{\oplus h^1(\End T)}
\oplus|\cO^6_1\ra_{0,-1/2}^{\oplus h^1(T)}~&&\rep{1}_0^{\oplus \{h^1(T)+h^1(T^\ast)+h^1(\End T)\}}\nonumber\\[1mm]
\end{align*}
\vspace{-15mm}

It is not hard to extend this analysis to a more general (0,2) CY NLSM with $\su(n)$ bundle $\cV \neq T_B$.  In particular, this offers certainly the most direct world-sheet perspective, in the spirit of~\cite{Distler:1987ee}, on marginal gauge-neutral deformations and agrees with spacetime~\cite{Donagi:2009ra,Anderson:2011ty} and world-sheet~\cite{Melnikov:2011ez} results on marginal deformations in the large radius limit.  This may be found in appendix~\ref{app:02NLSM}.

\subsubsection*{A hybrid example}
We will now illustrate how to set up the spectrum computation in a simple but non-trivial hybrid.
We consider the ``octic model''\footnote{The name comes from the large radius phase of this much-studied example.  Let $X_0$ be an octic hypersurface in the two-parameter toric resolution of the weighted projective space $\P^4_{\{2,2,2,1,1\}}$. The hybrid model 
$\cO(-2)\oplus\cO^{\oplus3}\rightarrow\P^1$ arises as one of the phases of the corresponding GLSM~{\protect \cite{Morrison:1994fr}}.}  with $B = \P^1$ and $X = \cO(-2)\oplus\cO^{\oplus3}$.  
The quantum numbers of the ground states of the twisted sectors, as well as charges of the fiber fields are given in table~\ref{tab:octicnums}.
\begin{table}[!t]
\renewcommand{\arraystretch}{1.3}
\begin{center}
\begin{tabular}{cc}
\begin{tabular}{|c|c|c|c|c|c|c|c|c|c|c|c|}
\hline
$k$ 		&$E_{|k\ra}$		&$\bq_{|k\ra}$		&$\bqb_{|k\ra}$   	&$\ell_k$ 	&$\nu_{i}$		&$\nut_{i}$	         &$\nu_I$			&$\nut_I$			\\ \hline
$0$		&$0$			&$-\ff{3}{2}$		&$-\ff{3}{2}$		&$0$		&$0$			&$0$  			&$0$	  		&$0$			\\ \hline
$1$		&$-1$			&$0$			&$-\ff{3}{2}$       	&$0$      		&$\ff{1}{8}$		&$-\ff{3}{8}$	 	&$0$			&$-\ff{1}{2}$		\\ \hline
$2$		&$0$			&$\ff{1}{2}$		&$-\ff{3}{2}$  		&$-2$		&$\ff{1}{4}$		&$-\ff{3}{4}$ 		&$0$			&$0$			\\ \hline
$3$		&$-\ff{1}{2}$		&$-1$			&$-\half$			&$0$		&$\ff{3}{8}$		&$-\ff{1}{8}$	 	&$0$			&$-\ff{1}{2}$		\\ \hline
$4$		&$0$			&$-\half$			&$-\half$			&$-2$		&$\ff{1}{2}$		&$-\ff{1}{2}$	 	&$0$			&$0$		    	\\ \hline
\end{tabular}
&
\begin{tabular}{|c|c|c|c|c|}
\hline
~ 		&$\phi^i$			&$\rho_i$				&$\chi^i$			&$\chib_i$  \\ \hline
$\bq$	&$\ff{1}{4}$		&$-\ff{1}{4}$			&$-\ff{3}{4}$			&$\ff{3}{4}$	\\ \hline
$\bqb$ &$\ff{1}{4}$		&$-\ff{1}{4}$			&$\ff{1}{4}$			&$-\ff{1}{4}$ \\ \hline
\end{tabular}
\end{tabular}
\end{center}
\caption{Quantum numbers for the octic model.}
\label{tab:octicnums}
\end{table}

In this example as well as those that follow $\Pic B = H^{2}(B,\Z)$, and the vacuum bundle $L_{|k\ra}$ is determined by a class in $H^2(B,\Z)$.  We label the class of the dual bundle $L_{|k\ra}^\ast$ by $\ell_k \in H^2(B,\Z)$.  In this example $\ell_k$ is simply the degree of the line bundle over $\P^1$.

Let us consider as an example the states at $E=0$ and $\bq=2$ in the $k=1$ sector. We see from \eqref{eq:E6decomposition} that these states  belong to  $\rep{1}_{2}$ of $\so(10)$.  Energy and charge considerations show that the relevant operators from~(\ref{eq:k1ops}) are $\cO^{1,s}$, and the states fit in a double complex
\begin{align}
\begin{xy}
\xymatrix@C=20mm@R=7mm{
  \Psi_{[2]}^{\alpha\beta}\chib_\alpha\chib_\beta|1\ra  & \Psi_{[5]}^\alpha\chib_\alpha |1\ra &0\\
\Psi^{\alpha\beta}_{[2]} \chib_\alpha\chib_\beta|1\ra
 & 
\Psi^{\alpha}_{[5]}\chib_\alpha |1\ra
&
 {\begin{matrix}
\Psi_{[8]}|1\ra
\end{matrix}}
}
\save="x"!LD+<-3mm,0pt>;"x"!RD+<0pt,0pt>**\dir{-}?>*\dir{>}\restore
\save="x"!LD+<75mm,-3mm>;"x"!LU+<75mm,2mm>**\dir{-}?>*\dir{>}\restore
\save!RD+<-82mm,-4mm>*{-\ff{3}{2}}\restore
\save!RD+<-42mm,-4mm>*{-\ff{1}{2}}\restore
\save!RD+<-7mm,-4mm>*{\ff{1}{2}}\restore
\save!RD+<3mm,-4mm>*{p}\restore
\save!CL+<72mm,16mm>*{U}\restore
\end{xy}
\end{align}
The wavefunctions satisfy $\cL_{V} \Psi^{\alpha\beta} = 0$ and $\cL_{V} \Psi^\alpha = \Psi^\alpha$; in practice this means that each $\Psi_{[d]}(y,\yb,\etab)$ is a quasi-homogeneous polynomial of degree $d$ in the fiber bosons $\phi^i$ if both indices are vertical, while it is of degree $d-1$ is one of the indices is horizontal.  
To limit clutter in the notation we suppressed the $\etab$s; their number is indicated by the $U$ grading.  Recall that the horizontal grading is by $p=\bq-u$.

Taking $\bQb_0$ cohomology at the relevant $\bq,\bqb, E$ eigenvalues indicated by the subscripts, we obtain
\begin{equation}\label{octicTY}
\begin{xy}
\xymatrix@C=10mm@R=10mm{
\left[ H^1(\bY, B_{2,0,0}) \right]_{2,-1/2,0}  & \left[ H^1(\bY, B_{1,0,0}) \right]_{2,1/2,0}   &0\\
\left[ H^0(\bY, B_{2,0,0} ) \right]_{2,-3/2,0}  
 & \left[ H^0(\bY, B_{1,0,0}) \right]_{2,-1/2,0}  
&
\left[ H^0(\bY, B_{0,0,0}) \right]_{2,1/2,0}  
}
\save="x"!LD+<-3mm,0pt>;"x"!RD+<0pt,0pt>**\dir{-}?>*\dir{>}\restore
\save="x"!LD+<98mm,-3mm>;"x"!LU+<98mm,2mm>**\dir{-}?>*\dir{>}\restore
\save!RD+<-120mm,-4mm>*{-\ff{3}{2}}\restore
\save!RD+<-70mm,-4mm>*{-\ff{1}{2}}\restore
\save!RD+<-20mm,-4mm>*{\ff{1}{2}}\restore
\save!RD+<0mm,-4mm>*{p}\restore
\save!CL+<95mm,16mm>*{U}\restore
\end{xy}
\end{equation}
To illustrate the counting, we concentrate on the dimension of 
\begin{align}
[H^0(\bY,B_{1,0,0})]_{2,-1/2,0}= [H^0(\bY,T_{\bY})]_{2,-1/2,0} = \bigoplus_{\sum_i r_i = 4} H^0_{\br}(\bY,T_{\bY})~.
\end{align}
The computation is simple since $\bY \simeq \bY' \times \C^3$, where $\bY'$ is the total space of $\cO(-2) \to \P^1$.  In this case, as we show in appendix~(\ref{app:sheaf}), the non-trivial graded cohomology groups are
\begin{align}
H^0_{r_1}(\bY',\cO_{\bY'}) = \C^{2r_1+1},\qquad
H^0_{r_1} (\bY',T_{\bY'}) = \C^{4r_1+4}~.
\end{align}
Decomposing $(T_{\bY})_{\br}$ according to~(\ref{eq:TSES}) we find two types of contributions to $H^0_{\br}(\bY,T_{\bY})$, those with $r_i \ge 0$, and those with $r_i=-1$ for $i=2,3,4$:
\begin{align}
 [H^0(\bY,T_{\bY})]_{2,-1/2,0} &= \bigoplus_{r_1=0}^4 \left[H^0_{r_1}(\bY',T_{\bY'}) \oplus H^0_{r_1} (\bY',\cO_{\bY'})^{\oplus 3}\right]\otimes \C^{\binom{6-r_1}{4-r_1}} \nonumber\\
 &\qquad\oplus\left[ \bigoplus_{r_1=0}^5 H^0_{r_1}(\bY',\cO_{\bY'}) \otimes \C^{6-r_1} \right]^{\oplus 3}~ \nonumber\\
 & = \C^{595}\oplus \C^{273} = \C^{868}.
\end{align} 
The factors of $\binom{6-r_1}{4-r_1}$  and $(6-r_1)$ arise from counting monomials, respectively, of degree $4-r_1$ in three variables and $5-r_1$ in two variables.

Computing the remaining cohomology groups in a similar fashion we obtain
the $E_1$ stage of the spectral sequence 
\begin{equation}\label{octicTYE1}
\begin{xy}
\xymatrix@C=10mm@R=10mm{
\C^{18}\ar[r]^{\bQb_W}  & \C^{21}  &0\\
\C^{126} \ar[r]^{\bQb_W}  
 & \C^{868}  \ar[r]^{\bQb_W}  
&
\C^{825}
}
\save="x"!LD+<-3mm,0pt>;"x"!RD+<10pt,0pt>**\dir{-}?>*\dir{>}\restore
\save="x"!LD+<37mm,-3mm>;"x"!LU+<37mm,5mm>**\dir{-}?>*\dir{>}\restore
\save!RD+<-45mm,-4mm>*{-\ff{3}{2}}\restore
\save!RD+<-26mm,-4mm>*{-\ff{1}{2}}\restore
\save!RD+<-4mm,-4mm>*{\ff{1}{2}}\restore
\save!RD+<3mm,-4mm>*{p}\restore
\save!CL+<33mm,14mm>*{U}\restore
\end{xy}
\end{equation}
Finally, we turn to the computation of the $\bQb_W$ cohomology for these states and for simplicity consider the Fermat superpotential
\begin{align}
W = S_{[8]}(\phi^1)^4+(\phi^2)^4+(\phi^3)^4+(\phi^4)^4~,
\end{align}
where $S_{[8]}\in  H^0(\P^1, \cO(8))$.  From~\eqref{eq:k1ops} we see that for the states appearing at $p=-\ff{3}{2}$
\begin{align}
\label{eq:QWp32}
\bQb_W \left( \Psi^{\alpha\beta}_{[2]u} \chib_\alpha\chib_\beta\right) |1\ra = 2 \Psi^{\alpha\beta}_{[2]} W_\beta \chib_\alpha|1\ra~,
\end{align}
and the derivatives of the superpotential that appear are ($a=2,3,4$)
\begin{align}
W_a &= 4(\phi^a)^3~, &W_1&=4S_{[8]}(\phi^1)^3~,    & W_I = \p_I S_{[8]} (\phi^1)^4~.
\end{align}
The map \eqref{eq:QWp32} has vanishing kernel, while the $\bQb_W$ action on the $p=-\ff{1}{2}$ states is
\begin{align}
\label{eq:QWp12}
\bQb_W \left( \Psi^\alpha_{[5]} \chib_\alpha \right)|1\ra = \Psi^\alpha_{[5]} W_\alpha ~.
\end{align}
Setting this to zero implies
$\Psi^\alpha_{[5]} = \Phi^{\alpha\beta}_{[2]}W_{\beta}$ for some
$ \Phi^{\alpha\beta}_{[2]}$ anti-symmetric in its indices. Hence the cohomology in the $(p,u)=(-\ff{1}{2},0)$ position is trivial, and the spectral sequence 
degenerates at
\begin{equation}\label{octicTYres}
\begin{xy}
\xymatrix@C=10mm@R=10mm{
0  & \C^3  &0\\
0
 & 0
&
\C^{83}
}
\save="x"!LD+<-3mm,0pt>;"x"!RD+<10pt,0pt>**\dir{-}?>*\dir{>}\restore
\save="x"!LD+<25mm,-3mm>;"x"!LU+<25mm,5mm>**\dir{-}?>*\dir{>}\restore
\save!RD+<-38mm,-4mm>*{-\ff{3}{2}}\restore
\save!RD+<-22mm,-4mm>*{-\ff{1}{2}}\restore
\save!RD+<-4mm,-4mm>*{\ff{1}{2}}\restore
\save!RD+<3mm,-4mm>*{p}\restore
\save!CL+<22mm,14mm>*{U}\restore
\end{xy}
\end{equation}
Here we count 86 anti-chiral states in the $\mathbf{1}_2$.   These correspond to the $83$ polynomial and the $3$ non-polynomial deformations of complex structure of the octic hypersurface now determined from the hybrid's point of view.

\subsubsection*{Extra states in $k=1$}

Multiplets with $q_i=\half$ can potentially give rise to additional massless states.  In a LGO theory these genuinely correspond to massive multiplets that can be integrated out without affecting the IR physics.  In general this is not so in the hybrid theory:  if a $q_i = \half$ field is non-trivially fibered then its mass vanishes on the discriminant of $W$ in $B$, and the field cannot be integrated out globally over $B$.  This leads to a rich structure entirely absent from LGO theories.  
 
To describe the additional operators with $h= 1$ we sadly need a little more notation.  Just in this section we use the indices $i', j'$, etc. to denote the multiplets with $q_{i'} = \half$; the $\alpha,\beta,\ldots$ continue to denote all the fields, while $I,J,\ldots$ denote the fields of the base geometry.  Let $X_{\half} \equiv \oplus_{i'} L_{i'}$ and $\cA$ be a holomorphic (in fact diagonal) connection on the bundle $X_{\half}\to B$.
The new operators are then
\begin{align}
\label{eq:massiveops1}
\cO^{7} &= \Psi^{7i'j'k'm'}(y^I) \chib_{i'}\chib_{j'}\chib_{k'}\chib_{m'}~,\qquad
\cO^{8} = ~:\Psi^{8 i'j'}_{I}(y^I) \chib_{i'} \chib_{j'} \chi^I :~,\nonumber\\[1.5mm]
\cO^{9} &=~ :\Psi^{9 i'j'} (y^I) (\rho_{i'} +\chi^I \cA^{k'}_{Ii'} \chib_{k'}) \chib_{j'}:~.
\end{align}
The wavefunctions are (0,u) forms valued in the following bundles:
\begin{align}
\Psi^{7} &:~  \wedge^4 X_{\half}~,&
\Psi^{8} &:~ \wedge^2 X_{\half} \otimes T_{B}^\ast~,&
\Psi^{9} &:~ X_{\half}\otimes X_{\half}~.
\end{align}
These operators have weight $h=1$ and charges
\begin{align}
\label{eq:massiveops2}
\xymatrix@R=1.5mm@C=2.5mm{
~ &
\cO^7_u&
\cO^8_u&
\cO^9_u
\\
\bq&
2&
0&
0
\\
\bqb&
u-2&
u-1&
u-1
}
\end{align}
The action of $\bQb_0$ on $\cO^7$ is simply to send $\Psi^7_u\to (-\pb \Psi^7)_{u+1}$.  
Since we used the holomorphic connection $\cA$ in $\cO^{9}$ to build a well-defined operator, the $\bQb_0$ action on $\cO^8_{u}+\cO^9_{u}$ is a bit more involved:
\begin{align}
\label{eq:massiveobs1}
\bQb_0 \cdot (\cO^8_u+\cO^9_u) & = - (\pb \Psi^{9i'j'})_{\Ib_0\cdots\Ib_u} \etab^{\Ib_0}\cdots\etab^{\Ib_u} (\rho_{i'} +\chi^I \cA^{k'}_{Ii'} \chib_{k'}) \chib_{j'} \nonumber\\[1.5mm]
&\qquad+ \left[ \obs(\Psi^{9})^{k'j'}_{I} - \pb \Psi^{8k'j'}_I\right]_{\Ib_0\cdots\Ib_u}\etab^{\Ib_0}\cdots\etab^{\Ib_u} \chib_{k'}\chib_{j'}\chi^I~,
\end{align}
where the linear map
\begin{align}
\label{eq:massiveobs2}
\obs : \Omega^{0,u}(X_{\half}\otimes X_{\half}) \to \Omega^{0,u+1}(\wedge^2 X_{\half}\otimes T^\ast_B)
\end{align}
is given by contracting $\Psi^9$ with the curvature $\cF = \pb \cA$~:
\begin{align}
\obs(\Psi^{9})^{k'j'}_{I~\Ib_0\cdots\Ib_u} d\yb^{\Ib_0} \cdots d\yb^{\Ib_u} \equiv
\ff{1}{2}\left( \cF_{I\Ib_0~ i'}^{~~~k'}\Psi^{9i'j'}_{\Ib_1\cdots \Ib_u} - \cF_{I\Ib_0~ i'}^{~~~j'}\Psi^{9i'k'}_{\Ib_1\cdots \Ib_u} \right) d\yb^{\Ib_0} \cdots d\yb^{\Ib_u}~.
\end{align}
It is easy to see that $\obs(\Psi^9)$ is $\pb$-closed when $\Psi^9$ is $\pb$-closed, so that $\cO^8_u+\cO^9_u$ is $\bQb_0$-closed iff $\pb \Psi^9 = 0$ and $\obs(\Psi^9)$ corresponds to the trivial class in $H^{u+1}(B, \wedge^2 X_{\half}\otimes T^\ast_B)$.  We will meet examples of such possible ``obstruction classes'' below, but for now we simply note that $\obs$ vanishes in a number of important cases that often arise in particular examples.  For instance, $\obs(\Psi^9_d)$ is clearly zero, and $\obs = 0$ for any $\Psi^{9} \in H^{\bullet}(B,L_{j'}\otimes L_{j'})$. A little less trivially, we can also show that $\obs$ vanishes for any $\Psi^{9} \in H^{\bullet}(B, \wedge^2 X_{\half}).$

The $\bQb_W$ action can also be determined;\footnote{ A little care is required in using point-splitting and the free OPE in computing the action on $\cO^9$.}  the results are:
\begin{align}
\bQb_W \cdot \cO^7 &= 4 W_{i'} \Psi^{7i'j'k'm'} \chib_{j'}\chib_{k'}\chib_{m'}~,\qquad
\bQb_W \cdot \cO^8 = 2 W_{i'} \Psi^{8i'j'}_{I}\chib_{j'} \chi^I~,\nonumber\\
\bQb_W \cdot \cO^9 &= \Psi^{9i'j'}\left[(\rho_{i'} -\cA^{i'}_{I}\chib_{i'} \chi^I) W_{j'} +(\chi^\alpha W_{i'\alpha} -\chi^I\cA^{k'}_{Ii'} W_{k'} )\chib_{j'}\right]~.
\end{align}

\subsection{$k>1$ (NS,R) sectors}
Finally, we turn to (NS,R) sectors with $1<k<2N-1$.  These sectors have, in general, two complications relative to the $k=1$ sector: in general $\bYk\neq \bY$, and $|k\ra$ may transform as a section of a nontrivial bundle over the base $B$.

\subsubsection*{The vacuum}

Recalling the discussion above~(\ref{eq:bYk}), we split the coordinates as $y^\alpha \to (y^I,\phi^{i'},\phi^A)$. The quantum numbers of the vacuum are then
write the vacuum energy as 
\begin{align} 
E_{|k\ra}  
&= -1 + \half\left[\sum_{i'} (\nu_{i'}-\frac{q_{i'}}{2}) + \sum_A (1-{\frac{q}{2}}-\nu_A)\right]~,\nonumber\\
\bq_{|k\ra}
&= \sum_{i'}(\frac{q_{i'}}{2}-\nu_{i'}) + \sum_A(1-\frac{q_A}{2}-\nu_A)~,\nonumber\\
\bqb_{|k\ra} 
&= \sum_{i'}(\frac{q_{i'}}{2}-\half-\nu_{i'}) + \sum_A(\nu_A-\frac{q_A}{2}+\half) - \frac{d}{2}~,
\end{align}
where $d$ is the dimension of the base $B$.  Note that in the twisted sectors $1<k<2N-1$ we have $E_{|k\ra} > -1$.
The vacuum bundle~(\ref{eq:bundlegroundst}) is given by 
\begin{align}\label{eq:vactrans}
L_{|k\ra} =  \otimes_A L_A^*\ .
\end{align}

\subsubsection*{Modes and transition functions}

Because we have $E_k>-1$ we can restrict attention to the subspace of the Hilbert space generated by the lowest modes of the left-moving fields.  That is, we truncate~(\ref{eq:modes}) to
\begin{align}\label{eq:moding}
y^\alpha(z) &= z^{\nu_\alpha-q_\alpha/2}(y^\alpha + z^{-1}\rho^{\dagger\alpha})~, &
\rho_\alpha(z) &= z^{q_\alpha/2-\nu_\alpha}(\rho_\alpha + z^{-1}y^\dagger_\alpha)~,\nonumber\\
\chi^\alpha(z) &= z^{\nut_\alpha-q_\alpha/2-\half}(\chi^\alpha + z^{-1}\chib^{\dagger\alpha})~, &
\chib_\alpha(z) &= z^{q_\alpha/2+\half-\nut_\alpha}(\chib_\alpha + z^{-1}\chi^\dagger_\alpha)\ ,
\end{align}
where in our restricted Hilbert space $\rho_I = 0$.

The transition functions for these oscillators follow by
expanding~(\ref{eq:patchfields}).  These show that 
$y^{\alpha'}$ are coordinates on $\bYk$, while $\chi^{\alpha'}$ ($\chib_{\alpha'}$) take values in $T_{\bYk}$ ($T^*_{\bYk}$).
On the other hand, $\phi^A$ and $\lambda^A$ take values in $\hat Z_k = \pi_k^*(\oplus L_A)$ and 
$\rho_A$ and $\lambdab_A$ in $\hat Z_k^*$.
As  is the case for $k=1$, $\rho_{i'}$ is not a covariant operator due to
the fermion bilinear term.

\subsubsection*{Conserved charges}

Inserting~(\ref{eq:moding}) into our expressions for the conserved charges~(\ref{eq:leftcharges}) we find in our Hilbert space
\begin{align}
L_0 &= \sum_\alpha \left[-\nu_\alpha\phi^\alpha\phi^\dagger_\alpha + 
(1-\nu_\alpha)\rho_\alpha\rho^{\dagger\alpha} + 
(1+\nut_\alpha)\chi^\alpha\chi^\dagger_\alpha -
\nut_\alpha \chib_\alpha\chib^{\dagger\alpha}\right]~, \nonumber\\
J_0 &= \sum_\alpha \left[(q_\alpha-1)(\chi^\alpha\chi^\dagger_\alpha - \chib_\alpha\chib^{\dagger\alpha})  - 
q_\alpha\left(y^\alpha y^\dagger_\alpha +
  \rho_\alpha\rho^{\dagger\alpha}\right)\right]~,\nonumber\\
\Jb_0 &= \sum_\alpha q_\alpha\left(-y^\alpha y^\dagger_\alpha - \rho_\alpha\rho^{\dagger\alpha} + \chi^\alpha\chi^\dagger_\alpha - \chib_\alpha\chib^{\dagger\alpha}\right)\ .\nonumber
\end{align}

\subsubsection*{States}

We again list the operators that can carry weight $h< 1$, suppressing
the right-moving $\etab^{\Ib}$ dependence.  These can contain at most
one operator with $h\ge\half$ and we organize them according to the
nature of this operator as
\begin{align}
\cO^{1,l,m} &= \Xi^{1lm}
{}^{\alpha',i'_2\cdots i'_l}_ {A_1\cdots A_m}
\chib_{\alpha'}\chib_{i'_2}\cdots\chib_{i'_l}\chi^{A_1}\cdots\chi^{A_m}\nonumber\\
\cO^{2,l,m} &= \Xi^{2lm}
{}^{i'_1\cdots i'_l}_ {\alpha',A_2\cdots A_m}
\chib_{i'_1}\cdots\chib_{i'_l}\chi^{\alpha'}\chi^{A_2}\cdots\chi^{A_m}\nonumber\\
\cO^{3,l,m} &= \Xi^{3lm}
{}^{B,i'_2\cdots i'_l}_ {A_1\cdots A_m}
\chib_B\chib_{i'_2}\cdots\chib_{i'_l}\chi^{A_1}\cdots\chi^{A_m}\\
\cO^{4,l,m} &= \Xi^{4lm}
{}^{i'_1\cdots i'_l}_ {B,A_1\cdots A_m}\phi^B
\chib_{i'_1}\cdots\chib_{i'_l}\chi^{A_1}\cdots\chi^{A_m}\nonumber\\
\cO^{5,l,m} &= \Xi^{5lm}
{}^{j',i'_1\cdots i'_l}_ {A_1\cdots A_m}
\left[\rho_{j'} - \cA_J{}^{k'}_{j'}\chi^J\chib_{k'}\right]
\chib_{i'_1}\cdots\chib_{i'_l}\chi^{A_1}\cdots\chi^{A_m}\nonumber\ .
\end{align}
In constructing $\cO^5$ we have introduced a holomorphic (and
diagonal) connection on $\oplus_{i'} L_{i'}$.  Here the $\Xi^t$
include the dependence on $y^{\alpha'}$ and $\rho_A$, as well as on the
right-moving zero modes of $\yb^{\Ib}$.  We can make this more explicit
by writing, for example, 
\begin{align}
\Xi^{1lm}{}^{\alpha',i'_2\cdots i'_l}_ {A_1\cdots A_m} = 
\sum_{\vec t} 
\Psi^{1lm}_{\vec t}(y){}^{\alpha',i'_2\cdots i'_l}_ {A_1\cdots A_m}
\prod_B \rho_B^{t_B+\sum_{a=1}^m \delta_{B,A_a}}\ ,
\end{align}
in terms of a vector of integers $t_B\ge -1$ such that no negative
powers of $\rho_B$ appear.   $\cO^1$ will now create
a well-defined state when acting on $|k\ra$ provided the wavefunction 
$\Psi^{1lm}_{\vec t}$ transforms as a section of a suitable bundle $\cE^{1lm}_{\vec
  t}$ over $\bYk$
\begin{align}
\cE^{1lm}_{\vec t} = 
T_{\bYk}\wedge\left(\wedge^{l-1}\pi_k^*(X_k)\right)
\otimes\left(\otimes_B(\pi_k^*L_B^{t_B+1})\right)\ .
\end{align}
Note that this takes into account the transformation properties of the
vacuum~(\ref{eq:vactrans}) and that the odd shift in the power of
$\rho_B$ is now seen to be sensible.   Incorporating the right-moving
fermion zero modes, the wavefunction is in general a (0,u) horizontal
form valued in this bundle.   These can be fine graded as
in~\ref{app:sheaf} by a vector of integers $\vec r = (r_{\alpha'})$.

Proceeding in an analogous way with the other operators we find that
the wavefunctions take values in the following bundles, organized by
$\vec t$ and the fine grading $\vec r$
\begin{align}
\cE^{1lm}_{\vec t,\vec r}(k) &= 
\left[T_{\bY_k}\wedge\left(\wedge^{l-1}\pi_k^*(X_k)\right)
\otimes\left(\otimes_A(\pi_k^*L_A^{t_A+1})\right)\right]_{\vec
r}\nonumber\\
\cE^{2lm}_{\vec t,\vec r}(k) &= 
\left[\left(\wedge^{l}\pi_k^*(X_k)\right)\otimes T^*_{\bY_k}
\otimes\left(\otimes_A(\pi_k^*L_A^{t_A+1})\right)\right]_{\vec r}\nonumber\\
\cE^{3lm}_{\vec t,\vec r}(k) &= 
\oplus_B\left[\left(\wedge^{l-1}\pi_k^*(X_k)\right)
\otimes\left(\otimes_A(\pi_k^*L_A^{t_A+1})\right)\right]_{\vec r}\\
\cE^{4lm}_{\vec t,\vec r}(k) &= 
\oplus_B\left[\left(\wedge^{l}\pi_k^*(X_k)\right)
\otimes\left(\otimes_A(\pi_k^*L_A^{t_A+1})\right)\right]_{\vec r}\nonumber\\
\cE^{5lm}_{\vec t,\vec r}(k) &= 
\left[\pi_k^*(X_k)\otimes \left(\wedge^{l}\pi_k^*(X_k)\right)
\otimes\left(\otimes_A(\pi_k^*L_A^{t_A+1})\right)\right]_{\vec r}\nonumber
\end{align}
We need to consider all $\vec t,\vec r$ that contain states $\cO|k\ra$ with
$E=0$.

\subsubsection*{$\bQb$ and cohomology}

On states of the form $\cO^1_u|k\ra,\ldots, \cO^4_u|k\ra$ 
$\bQb_0$ acts as $-\pb$ on horizontal (0,u) forms valued in holomorphic bundles over $\bYk$,
and $\bQb_0$ cohomology is the horizontal Dolbeault cohomology.
The action on states of the form $\cO^5|k\ra$ has an added term of the sort already familiar from (\ref{eq:massiveobs1},\ref{eq:massiveobs2}) for the ``massive'' states in the $k=1$ sector:
\begin{align}\label{eq:QbPtwo}
\bQb_0 \cO_u^5|k\ra = -\etab^{\Kb}\left[\pb_{\Kb} \cO_u^5{}^{j'}+
\cF_{\Kb J}{}^{j'}_{k'}\chi^J\chib_{j'} (\Xi^{5k'}_{u})^{i'_1\cdots i'_l}_{A_1\cdots A_m}\chib_{i'_1}\cdots\chib_{i'_l}\chi^{A_1}\cdots\chi^{A_m} \right]|k\ra\ ,
\end{align}
where $\cF$ is the curvature of $\cA$.  For $\pb$-closed $\Psi^5$, the
additional ``obstruction'' term is $\pb$-closed and gives a linear map
\begin{align}
\obs : \Omega^{0,u}(\cE^{5l,m}) \to \Omega^{0,u+1}(\cE^{4(l+1),m}\otimes\pi^\ast_k T_B^\ast)~.
\end{align}
If $\obs(\Psi^5)$ is exact, then we can construct a $\bQb_0$-closed state just as we saw in the $k=1$ case.
We have not encountered a nontrivial obstruction term in any of the
examples we considered, and  in~\ref{subsec:cpt} we argue that this will be
the case in any well-defined model.

The action of $\bQb_W$ is given by the mode expansion of 
\begin{align}
\bQb_W = \oint~\frac{dz}{2\pi i} \chi^\alpha W_\alpha = \chi^\alpha\oint ~\frac{dz}{2\pi i}
z^{\nut_\alpha-q_\alpha/2-1/2} W_\alpha + \chib^{\dagger\alpha}\oint~ \frac{dz}{2\pi i}
z^{\nut_\alpha-q_\alpha/2-3/2} W_\alpha \ ,
\end{align}
where we write
\begin{align}
W_\alpha =
W_\alpha\left(z^{\nu_\beta-q_\beta/2}(\phi^\beta +
    z^{-1}\rho^{\dagger\beta})\right)\ .
\end{align}
We can use the homogeneity relation
$W_\alpha(\lambda^q\phi^\beta) = \lambda^{1-q_\alpha}
W_\alpha(\phi^\beta)$ and simplify this to
\begin{align}\label{eq:Wmodes}
\bQb_W = \chi^\alpha\oint \frac{dz}{2\pi i}
z^{\nut_\alpha} W_\alpha\left(z^{\nu_\beta}(\phi^\beta +
    z^{-1}\rho^{\dagger\beta})\right) + 
\chib^{\dagger\alpha}\oint \frac{dz}{2\pi i}
z^{\nut_\alpha-1} W_\alpha\left(z^{\nu_\beta}(\phi^\beta +
    z^{-1}\rho^{\dagger\beta})\right) \ .
\end{align}

\subsection{ Comments on CPT}
\label{subsec:cpt}

The spectrum we obtain should be invariant under CPT.  This means that
for any massless state with charge $(\bq,\bqb)$ in the $k$ sector we
should find a massless state with charge $(-\bq,-\bqb)$ in the $2N-k$
sector.  In this section we will discuss how this works for sectors with 
odd $k$.  To avoid additional notational elaborations we will make the simplifying
assumption that $\nut < 0$ for all fields.\footnote{When this is not the case there
are, as in the (R,R) sectors, $\chi$ and $\chib$ zero-modes.  It should be possible
to extend the CPT discussion to these situations as well.}  As we will now argue,
CPT invariance essentially reduces to Serre duality for Dolbeault cohomology on $B$, as well
as a natural dual action of $\bQb_W$.

\subsubsection*{A pairing on the Hilbert spaces} 
The two-point function in the CFT is a
natural pairing between the conjugate sectors respecting charge
conservation and pairing states with the same energy, and given the quantum orbifold symmetry
we expect that the Hilbert spaces of states in the $|k\ra$ and $|2N-k\ra$ sectors are dual to each other in
this way.

From the expressions above it is clear that the vacua satisfy
\begin{align}
E_{|2N-k\ra} = E_{|k\ra};\qquad (\bq_{|2N-k\ra},\bqb_{|2N-k\ra}) = (-\bq_{|k\ra},d-\bqb_{|k\ra})
\end{align}
while the moding in the conjugate sectors is related by 
\begin{align}\label{eq:conjnus}
\nu_\alpha\leftrightarrow 1-\nu_\alpha;\qquad \nut_\alpha\leftrightarrow
-1-\nut_\alpha\ .
\end{align}
This implies that the fields $\phi^{i'}$ for which $\tau=1/2$ in the
$k$ sector have $\tau=3/2$ in the conjugate $2N-k$ sector, and vice
versa, so that we have 
\begin{align}
\bYk &= {\rm tot} (\oplus_{i'} L_{i'}{\rightarrow} B)\qquad &L_{|k\ra} &=
\otimes_A L_A^*\nonumber\\
\bY_{\!\!2N-k} &={\rm tot} (\oplus L_A{\rightarrow} B)\qquad &L_{|2N-k\ra} &=
\otimes_{i'} L_{i'}^*~.
\end{align}
In particular $L_{|k\ra}\otimes L_{|2N-k\ra} = K^\ast_B$.   For any state with
weight $h$ and charge $(\bq,\bqb)$ in the $k$ sector, we can find a
state with the same weight and charge $(-\bq,d-\bqb)$ in the $2N-k$
sector by exchanging the oscillator excitations according to 
\begin{align}
y^\alpha\leftrightarrow\rho_\alpha\qquad\chi^\alpha\leftrightarrow\chib_\alpha\ .
\end{align}
This is enough to show that at the level of left-moving oscillators the two-point
function leads to a pairing between the state spaces defined above,
which respects $\bq$ and violates $\bqb$ by $d$.   If we denote 
$\cH^{tlm}_{\vec t,\vec r}(k) = \Gamma(\cE^{tlm}_{\vec t,\vec r}(k))$, 
then the pairing takes the form 
\begin{align}\label{eq:pairs}
\cH^{1\oplus 2\oplus 3}{}^{lm}_{\vec t,\vec r}(k)\times 
\cH^{1\oplus 2\oplus 3}{}^{ml}_{\vec r,\vec t}(2N-k)
&\to \C\nonumber\\
\cH^{4lm}_{\vec t,\vec r}(k)\times\cH^{5ml}_{\vec r,\vec t}(2N-k) &\to
\C\ ,
\end{align}

\subsubsection*{$\bQb_0$ and Serre duality}
The pairing descends to $\bQb_0$ cohomology, and in a
reasonable physical theory this must be nondegenerate.   This will be
the case if  
\begin{align}
H^\bullet_{\bQb_0}\left(\cH^{lm}_k(\vec t,\vec r)\right) =
\left[H^{d-\bullet}_{\bQb_0}\left(\cH^{ml}_{2N-k}(\vec r,\vec t)\right)\right]^*\ .
\end{align}
For the first line in~(\ref{eq:pairs}), in which $\bQb_0$ acts as
$-\pb$,  this is in fact equivalent to Serre duality.   For
simplicity let's see first how this works in $\cH^{111}_{\vec r,\vec t}$.  The
fine grading on $H^\bullet(T_{\bY_k})$ can be obtained from the long exact sequence (LES)
following from the short exact sequence (SES)~(\ref{eq:Tsections}) 
\begin{align}
\xymatrix{0 \ar[r]& \oplus_i (\pi_k^\ast L_i)_{\br+\bx_i} \ar[r] &
  (T_{\bYk})_{\br} \ar[r] &(\pi_k^\ast T_B)_{\br} \ar[r] & 0}\ ,
\end{align}
which we here encounter twisted by a vector bundle (so still exact) as
\begin{align}\label{eq:sestwst}
\xymatrix{0 \ar[r]& \oplus_i (\pi_k^\ast
  L_i)_{\br+\bx_i}\otimes\Vh_{\vec t} \ar[r] &
  (T_{\bYk})_{\br}\otimes\Vh_{\vec t} \ar[r] &(\pi_k^\ast T_B)_{\br} \otimes\Vh_{\vec t}\ar[r] & 0}\ ,
\end{align}
where
\begin{align}
\Vh_{\vec t} = 
\pi_k^*\left[\oplus_B\left(\otimes_A(L_A^{t_A+1})\right)\right]\ .
\end{align}
The bundles on either end of the SES are pulled back from $B$, and we
can use~(\ref{eq:sheafpullback}) to compute their cohomology.  Thus
\begin{align}
H^\bullet_{\vec r}\left(\bYk,\oplus_i (\pi_k^\ast
  L_i)_{\br+\bx_i}\otimes\Vh_{\vec t}\right)
 = 
H^\bullet\left(B,\oplus_{i,B}\left(
\otimes_A(L_A^{t_A+1})
\otimes\left(\otimes_j (L_j^*)^{r_j}
\right)\right)\right)\ ,
\end{align}
while
\begin{align}
H^\bullet_{\vec r}\left(\bYk,(\pi_k^\ast\, T_B)_{\br} \otimes\Vh_{\vec
    t}\right) = 
H^\bullet\left(B,T_B\otimes\left(\oplus_B\left(\otimes_A(L_A^{t_A+1})
\otimes \left(\otimes_j (L_j^*)^{r_j}\right)\right) \right)\right)\ .
\end{align}
Recalling that $K_B = \otimes_\alpha L_\alpha$, these are Serre dual, respectively, to 
\begin{align}
&H^{d-\bullet}\left(B,\oplus_{i,B}\left(
\otimes_A(L_A^*)^{t_A}
\otimes\left(\otimes_j (L_j)^{r_j+1}
\right)\right)\right)\nonumber\\
&= H^{d-\bullet}\left(\bY_{\!\!2N-k},\oplus_{i,B}\left(\pi_{2N-k}^*(L_A^*)_{\vec
      t - \vec y_A}\otimes\left(\otimes_j(\hat L_j^{r_j+1})\right)\right)\right)
\end{align}
and 
\begin{align}
&H^{d-\bullet}\left(B,T^*_B\otimes\left(\oplus_B\left(\otimes_A(L_A^*)^{t_A}
\otimes \left(\otimes_j (L_j^{r_j+1})\right)\right)
\right)\right)\nonumber\\
&=  H^{d-\bullet}\left(\bY_{\!\!2N-k},\left(\pi^*_{2N-k}\,T^*_B\right)_{\vec t}\otimes\left(
\oplus_j(\pi_{2N-k}^*(L_j)^{r_j+1})\right)\right)\ .
\end{align}
Inserting this result into the dual LES we find 
\begin{align}
H^\bullet( (T_{\bYk})_{\br}\otimes\Vh_{\vec t}) = 
\left[H^{d-\bullet}( (T_{\bY_{\!\!2N-k}})_{\vec t}\otimes\Vh_{\vec r})\right]^*
\end{align}
with a suitable natural definition for $\Vh_{\vec r}$.

Higher powers of the tangent/cotangent bundles are fine graded by
recursively using the same SES and the dual, so recursively applying
this argument we find that Serre duality implies CPT in the sense
above whenever we can use $\bQb_0 = -\pb$.  This argument will fail if
nontrivial obstruction classes arise in~(\ref{eq:QbPtwo}), because
no such obstruction can arise for the dual states in $\cH^4$.  We conclude that in
reasonable physical theories there will be no nontrivial obstructions
in the twisted sectors.

\subsubsection*{$\bQb_W$ and CPT}
Given that the cohomology of
$\bQb_0$ produces a spectrum consistent with CPT, we can also show
that the action of $\bQb_W$ is consistent with this.   Consider a monomial
in $W_\alpha$ that contributes to $\bQb_W$ in the $k$ sector a term
\begin{align}
\chi^\alpha\prod_\beta\left[(\phi^\beta)^{m_\beta}
  (\rho^{\dagger\beta})^{n_\beta}\right]\ .
\end{align}
This means that 
\begin{align}
\sum_\beta\left[\nu_{\beta,k}(m_\beta+n_\beta) - n_\beta\right] =
 - \nut_{\alpha,k} -1\ .
\end{align}
Using~(\ref{eq:conjnus}) we see that this implies
\begin{align}
\sum_\beta\left[\nu_{\beta,2N-k}(m_\beta + n_\beta) - m_\beta\right]
= \sum_\beta\left[-\nu_{\beta,k}(m_\beta + n_\beta)+n_\beta \right] = 
\nut_{\alpha,k}+1 = -\nut_{\alpha,2N-k}\ ,
\end{align}
which means that the same monomial contributes a term 
\begin{align}
\chib^{\dagger\alpha}\prod_\beta\left[(\phi^\beta)^{n_\beta}
  (\rho^{\dagger\beta})^{m_\beta}\right]
\end{align}
to $\bQb_W$ in the $2N-k$ sector.   This acts in precisely the
appropriately dual way on the states as mapped above, showing that CPT
is maintained as a symmetry after taking $\bQb_W$ cohomology.


\section{Examples}\label{s:Examples}

In this section we will apply the techniques developed in the previous sections to a number of hybrid examples.  In each case we will focus on characterizing first order deformations that preserve (0,2) superconformal invariance and the $\Le_8\oplus \Le_6$ spacetime gauge symmetry.  

The infinitesimal deformations which preserve (2,2) symmetry parametrize the tangent space of the (2,2) moduli space.  They are not obstructed and in a large radius limit are identified with complex structure and complexified K\"ahler moduli of the CY. There is a well-known correspondence between the (2,2) moduli and the $\Le_6$-charged matter, and we will borrow the large radius notation by denoting the number of chiral $\mathbf{27}$'s and $\mathbf{\overline{27}}$'s in the hybrid computation by $h^{1,1}$ and $h^{2,1}$ respectively.

More interesting are the deformations which  only preserve (0,2) superconformal invariance. 
The computation of the number of massless gauge singlets associated to these deformations, which we indicate as $\mathcal{M}$,
is the main goal of this section.  These singlets arise in (NS,R), i.e. the odd $k$ sectors.
In the following we will compute $\cM$ in three examples that illustrate a number of technical and conceptual points.
\begin{enumerate}
\item
For the first example we choose the simplest possible base, i.e. $B=\P^1$.  This is a good warm-up for more difficult cases and is of interest in its own right since the model can be found as a phase of a GLSM without a large radius limit in its K\"ahler moduli space.
In fact, it can be shown~\cite{Aspinwall:1994cf} that $h^{1,1}=1$, and the only other phase is a LGO.
\item
In the second example we describe a model in the broader orbi-bundle set-up with $B=\P^3$.
It will be clear that most of our discussion above was restricted to the case in which $X$ is a sum of line bundles solely for ease of exposition.
This example also give us a chance to compute a higher order differential (it will turn out to be zero).
\item
In the last example we consider the case in which one of the line bundles defining $X$ is positive, and $B=\mathbb{F}_0$ is not a projective space.
\end{enumerate}

While our construction does not depend on a GLSM embedding,
all of these models do arise as phases of a GLSM.
That gives us the possibility to compare the hybrid spectrum with the spectrum known in other phases.
What we discover is that while in the hybrid limit extra singlets appear at a particular complex structure or K\"ahler form, 
there is no evidence of world-sheet instanton corrections to masses of $\Le_6$ singlets.

\subsection{A hybrid with no large radius}

We begin with the model $X=\cO(-2)\oplus\cO^{\oplus4}$ and $B=\P^1$ with superpotential
\begin{align}
\label{genWnoLR}
W =  \sum_{p=0}^2 F_{[2p]}(\phi^1)^p. 
\end{align}
Some notational clarifications are in order: it is convenient to distinguish between the trivial and non-trivial fiber indices, so let $a,b=2,\dots,5$;  moreover, let $F_{[d]}$ be a generic polynomial of degree $4-d$ in the $\phi^a$'s, whose coefficients belong to $H^0(\P^1,\cO(d))$. 
The left- and right-moving charges for the fields and the quantum numbers of the twisted ground states are summarized in table \ref{table:noLR}.

The orbifold action $\Gamma=\Z_8$ introduces $7$ twisted sectors;  because of CPT invariance to compute the number of massless $\Le_6$-singlets it is sufficient to study the $k=1$ and $k=3$ sectors.

\subsubsection*{$k=1$ sector}
\begin{table}[!t]
\renewcommand{\arraystretch}{1.3}
\begin{center}
\begin{tabular}{cc}
\begin{tabular}{|c|c|c|c|c|c|c|c|c|c|}
\hline
$k$ 		&$E_{|k\ra}$	&$\bq_{|k\ra}$	&$\bqb_{|k\ra}$ 	&$\ell_k$			&$\nu_a,\nu_1$		&$\nut_a,\nut_1$					
\\ \hline
$0$		&$0$		&$-\ff{3}{2}$	&$-\ff{3}{2}$		&$0$			&$0,0$				&$0,0$		
\\ \hline
$1$		&$-1$		&$0$		&$-\ff{3}{2}$		&$0$			&$\ff{1}{8},\ff{1}{4}$		&$-\ff{3}{8},-\ff{1}{4}$
\\ \hline
$2$		&$0$		&$\half$		&$-\ff{3}{2}$		&$-2$			&$\ff{1}{4},\ff{1}{2}$		&$-\ff{3}{4},-\ff{1}{2}$				
\\ \hline
$3$		&$-\ff{1}{2}$	&$-1$		&$-\half$			&$-2$			&$\ff{3}{8},\ff{3}{4}$		&$-\ff{1}{8},-\ff{3}{4}$				
\\ \hline
$4$		&$0$		&$-\half$		&$-\half$			&$0$			&$\ff{1}{2},0$			&$-\half,0$		
\\ \hline
\end{tabular}&
\begin{tabular}{|c|c|c|c|c|}
\hline
~ 		&$\phi^i,\phi^1$			&$\rho_i,\rho_1$				&$\chi^i,\chi^1$			&$\chib_i,\chib_1$  		\\ \hline
$\bq$	&$\ff{1}{4},\half$			&$-\ff{1}{4},-\half$				&$-\ff{3}{4},-\ff{1}{2}$			&$\ff{3}{4},\half$	\\ \hline
$\bqb$      &$\ff{1}{4},\half$			&$-\ff{1}{4},-\half$				&$\ff{1}{4},\ff{1}{2}$			&$-\ff{1}{4},-\half$ \\ \hline
\end{tabular}
\end{tabular}
\end{center}
\caption{Quantum numbers for the $X=\cO(-2)\oplus\cO^{\oplus4}\rightarrow \P^1$ model.}
\label{table:noLR}
\end{table}
The $E_1$ stage of the spectral sequence is obtained by taking $H_{\bQb_0}(\cH)$ as described in section \ref{ss:massless}
and we reproduce here the result, where the subscripts denote the dimension of the respective cohomology groups
\begin{equation}\label{eq:spectrseqE1noLRk1}
\begin{matrix}\vspace{20mm}\\E_1^{p,u}:\end{matrix}
\begin{xy}
\xymatrix@C=15mm@R=5mm{
H^1\left(\bY,B_{1,0,0}\right)_{3} 
   \ar[r]^{\bQb_W}&
{\begin{matrix}
H^1\left(\bY,B_{0,0,1} \right)_{10}
\\\oplus\\ 
H^1\left(\bY, B_{1,1,0} \right)_{63}
\end{matrix}}   \ar[r]^{\bQb_W}  &
H^1\left(\bY, B_{0,1,0} \right)_{35}   \\
   &
{\begin{matrix}
H^0\left(\bY,B_{0,0,1}\right)_{20}
\\\oplus\\ 
H^0\left(\bY,B_{1,1,0} \right)_{17}
\end{matrix}}   \ar[r]^{\bQb_W}  
  &
{\begin{matrix}
H^0\left(\bY, B_{0,1,0}\right)_{176}
\end{matrix}}
}
\save="x"!LD+<-6mm,0pt>;"x"!RD+<40pt,0pt>**\dir{-}?>*\dir{>}\restore
\save="x"!LD+<120mm,-3mm>;"x"!LU+<120mm,2mm>**\dir{-}?>*\dir{>}\restore
\save!RD+<-62mm,-4mm>*{-\ff{3}{2}}\restore
\save!RD+<-105mm,-4mm>*{-\ff{5}{2}}\restore
\save!RD+<-18mm,-4mm>*{-\half}\restore
\save!RD+<13mm,-3mm>*{p}\restore
\save!CL+<117mm,21mm>*{U}\restore
\end{xy}
\end{equation}

The lowest row of the sequence provides an example of the universal structure of currents we indicated above in~(\ref{eq:currents}), and for generic $W$ the kernel is one-dimensional, corresponding to the $\GUL$ symmetry.  By choosing a particular form of the superpotential (\ref{genWnoLR}) we can increase $\ker \bQb_W$, and the additional vectors 
correspond to an enhanced symmetry at the special locus in the moduli space.

In order to compute the cohomology of the top row of \eqref{eq:spectrseqE1noLRk1} let us list all the states contributing at $E_1^{p,1}$:
\begin{equation}\label{eq:k1cohomnoLR}
\begin{xy}
\xymatrix@C=20mm@R=10mm{
V \rho_1 \chib_1|1\ra_{3} 
   \ar[r]^-{\bQb_W} &
  { \begin{matrix}
  H_{[2]} \chib_1 \chi^I |1\ra_{30}  \\\oplus\\
G_{[1]} \chib_b \chi^I |1\ra_{16}  \\\oplus\\
G_{[1]} \chib_1 \chi^b |1\ra_{16} \\\oplus\\
 \Phi_I \phi^1 \chib_1 \chi^I |1\ra_{1}  \\\oplus\\
  G_{[2]} \rho_1|1\ra_{10}  \\\oplus\\
\Psi_I \pz y^I |1\ra_1
 \end{matrix} }
   \ar[r]^-{\bQb_W} &
G_{[4]} \chi^I |1\ra_{35}
}
\end{xy}
\end{equation}
where $G_{[d]}$ and $H_{[d]}$ are generic polynomials of degree $d$ in the $\phi^a$'s with coefficients in $H^1(\P^1,\cO(-2))$ and  $H^1(\P^1,\cO(-4))$, respectively, while $\Psi_I, \Phi_I \in H^1(\P^1,\cO(-2)) $. 
First, consider the map on the left. We have the state $V \rho_1 \chib_1|1\ra$ where $V \in H^1(\bY, T_{\bY}\otimes T_{\bY})\simeq H^1\left(\P^1, \cO(-4) \right) $. Under $\bQb_W$ it maps to
\begin{align}
\label{eq:massiveP1noLR}
\bQb_W V \rho_1 \chib_1|1\ra &= V \left( \p_1 W \rho_1+ \p_{11}W \chi^1 \chib_1 + \p_{1I}W \chi^I  \chib_1 \right) |1\ra + V\left( \p_{1a}W \chi^a\chib_1  \right)|1\ra ~.
\end{align}
Since $\p_1W, \p_{1a}W \in \Gamma(\P^1,\cO(2))$, it follows that $V\p_1W, V\p_{1a}W \in H^1 \left( \P^1, \cO(-2)\right)$.
To compute the dimension of the cokernel of this map we first note that if we restrict the superpotential to its Fermat form, namely $W=  \sum_{i=2}^5 (\phi^i)^4 + S_{[4]}(\phi^1)^2$, we have
\begin{align}
\bQb_W V \rho_1 \chib_1|1\ra &= 2V \left( \phi^1 S_{[4]} \rho_1+ S_{[4]} \chi^1 \chib_1 +  \phi^1 \p_IS_{[4]}\chi^I  \chib_1 \right) |1\ra ~.
\end{align}
Since $V \phi^1 S_{[4]} \in H^1 \left( \P^1, \cO(-2)\right)$ and $h^1(\P^1, \cO(-2))=1$ the kernel at Fermat is 2-dimensional. 

Adding to $W$ a term of the form $S_{[2]}\phi^2\phi^3\phi^1+ T_{[2]}\phi^4\phi^5\phi^1$, where $S_{[2]}, T_{[2]}\in \Gamma(B,\cO(2))$, we find that \eqref{eq:massiveP1noLR} reads
\begin{align}
\bQb_W V \rho_1 \chib_1|1\ra &= \underbrace{V \left( \p_1W \rho_1+ \p_{11}W \chi^1 \chib_1 + \p_{1I}W \chi^I  \chib_1 \right) |1\ra}_{\mbox{Fermat}} \nonumber\\
& \quad + V S_{[2]}\left(  \phi^2 \chi^3 + \phi^3\chi^2\right) \chib_1  |1\ra + V T_{[2]}\left(  \phi^4 \chi^5 + \phi^5\chi^4\right) \chib_1  |1\ra ~,
\end{align}
and the map is injective for $W$ generic enough.
Now, for the map on the right in \eqref{eq:k1cohomnoLR} we have
\begin{align}
\bQb_W \left( \Psi_{ab}\rho_1 +\Psi_{ab,I} \chi^I\chib_1 \right)\phi^a\phi^b |1\ra &= \p_\alpha\left( \Psi_{ab} \p_1 W\right) \chi^\alpha \phi^a\phi^b |1\ra \nonumber\\
\bQb_W \Sigma_{abI}\chi^I\chib_1 \phi^a\phi^b|1\ra  & = -\Sigma_{abI}\chi^I \p_1 W \phi^a\phi^b |1\ra \nonumber\\
\bQb_W V_a^b \phi^a \chi^I \chib_b |1\ra &= -V_a^b \phi^a \p_b W \chi^I |1\ra \nonumber\\
\bQb_W \Phi_I \phi^1 \chi^I \chib_1 |1\ra &= -\Phi_I \phi^1 \p_1 W \chi^I |1\ra 
\end{align}
The cokernel of this map is thus any object of the form $\Psi_{abcd} \phi^a\phi^b\phi^c\phi^d \chi^I |1\ra$ for $\Psi_{abcd}\in H^1(\P^1,\cO(-2))$, which cannot be written as
$\p_1 W \phi^a\phi^b \chi^I |1\ra$ or $\phi^a\p_b W \chi^I |1\ra$.  We find a $9$-dimensional space. 
Thus, the $E_2$ stage of the spectral sequence is
\begin{equation}
\begin{matrix}\vspace{10mm}\\E_2^{p,u}:\end{matrix}
\begin{xy}
\xymatrix@C=8mm@R=8mm{\;\;\;
0 &
\C^{45}
  &
\C^{9}  \\ \;\;\;
 &
\C^1  &
\C^{139}
}
\save="x"!LD+<3mm,0pt>;"x"!RD+<40pt,0pt>**\dir{-}?>*\dir{>}\restore
\save="x"!LD+<45mm,-3mm>;"x"!LU+<45mm,2mm>**\dir{-}?>*\dir{>}\restore
\save!RD+<-36mm,-4mm>*{-\ff{5}{2}}\restore
\save!RD+<-23mm,-4mm>*{-\ff{3}{2}}\restore
\save!RD+<-6mm,-4mm>*{-\ff{1}{2}}\restore
\save!RD+<12mm,-4mm>*{p}\restore
\save!CL+<42mm,12mm>*{U}\restore
\end{xy}
\end{equation}
and obviously all higher differentials vanish. Hence the spectral sequence degenerates already at this stage, $E_\infty = E_2$.
Thus, in this sector we count $45+139=184$ chiral  and $9$ anti-chiral $\Le_6$-singlets.

\subsubsection*{$k=3$ sector}
The $k=3$ ground state has a non-trivial vacuum bundle $L_{|3\ra} = \cO(2)$ and, as discussed in section \ref{s:twisted}, we must distinguish between light and heavy fields. 
In particular we have $A=1$, $i' = 2,\dots,5$, $\alpha' = (I,i')$, while the geometry is determined by
$\bY_{\!\!3}$, the total space of $\cO^{\oplus4}\xrightarrow{\pi_3}\P^1$.
The expansion of $\bQb_W$ in this sector takes the form
\begin{align}\label{QbWk3noLR}
\bQb_W = \chib^A{}^\dag \p_A W + \chi^A \p_{Ai}W \rho^i{}^\dag + \chib^{\alpha'}{}^\dag \p_{\alpha' i} W \rho^i{}^\dag + \chi^{\alpha'} \p_{\alpha' ij}W \rho^i{}^\dag\rho^j{}^\dag~.
\end{align}
The $E_1$ stage of the spectral sequence is given by
 \begin{equation}\label{E1k3noLR}
\begin{matrix}\vspace{10mm}\\E_1^{p,u}:\end{matrix}
\begin{xy}
\xymatrix@C=12mm@R=10mm{
\;\;\; H^1(\bY_{\!\!3}, \pi_3^\ast\left( L_6^\ast \right)\otimes \wedge^2 T_{\bY_{\!\!3}} )_{18}  \ar[r]^-{\bQb_W} & 
H^1(\bY_{\!\!3}, T_{\bY_{\!\!3}}  )_{16}   \\
\;\;\; 0 & 
{\begin{matrix}
H^0(\bY_{\!\!3}, \wedge^2 T_{\bY_3} \otimes T^\ast_{\bY_3})_{6} \\\oplus\\
H^0(\bY_{\!\!3},T_{\bY_3} )_1
\end{matrix}}
}
\save="x"!LD+<1mm,0pt>;"x"!RD+<33pt,0pt>**\dir{-}?>*\dir{>}\restore
\save="x"!LD+<107mm,-3mm>;"x"!LU+<107mm,8mm>**\dir{-}?>*\dir{>}\restore
\save!RD+<-76mm,-4mm>*{-\ff{3}{2}}\restore
\save!RD+<-20mm,-4mm>*{-\half}\restore
\save!RD+<11mm,-3mm>*{p}\restore
\save!CL+<104mm,23mm>*{U}\restore
\end{xy}
\end{equation}
Now, the only non-trivial map is at $U=1$, where
\begin{align}
\label{eq:bqbWk3noLR}
\bQb_W V^{ab}\rho_1\chib_a\chib_b |3\ra = 2V^{ab} \rho_1\p_{1a}W \chib_b |3\ra \neq 0 ~.
\end{align}
 The RHS never vanishes, giving a $6$-dimensional image. Hence, the spectral sequence degenerates at the $E_2$ term
\begin{equation}
\begin{matrix}\vspace{8mm}\\E_2^{p,u}:\end{matrix}
\begin{xy}
\xymatrix@C=8mm@R=8mm{\;\;\;
\C^{12}
  &
\C^{10}  \\ 
0  &
\C^{7}
}
\save="x"!LD+<3mm,0pt>;"x"!RD+<30pt,0pt>**\dir{-}?>*\dir{>}\restore
\save="x"!LD+<32mm,-3mm>;"x"!LU+<32mm,2mm>**\dir{-}?>*\dir{>}\restore
\save!RD+<-23mm,-4mm>*{-\ff{3}{2}}\restore
\save!RD+<-6mm,-4mm>*{-\ff{1}{2}}\restore
\save!RD+<10mm,-3mm>*{p}\restore
\save!CL+<29mm,12mm>*{U}\restore
\end{xy}
\end{equation}
Hence we count 19 chiral and 10 anti-chiral states for a total of $222$ $\Le_6$ singlets.
By similar methods we compute $h^{2,1}=61$ and $h^{1,1}=1$, yielding $\cM=160$.

\subsection{The orbi-bundle}

Now we present an example in which $X$ is not a sum of line bundles, but a more general orbi-bundle.
Let us take $B=\P^3$ and $X=\cO(-5/2)\oplus\cO(-3/2)$ along with the quasi-homogeneous superpotential
\begin{align}
W= S_5 (\phi^1)^2 + S_4\phi^1\phi^2+ S_3 (\phi^2)^2 , 
\end{align}
where $S_d\in H^0(B,\cO(d))$.
The ground state quantum numbers and charges of the fields  are given in table \ref{table:orbinums}, and to find the singlets we need only consider the first twisted sector.  

The first stage of the spectral sequence is
\begin{equation}
\begin{matrix}\vspace{20mm}\\E_1^{p,u}:\end{matrix}
\begin{xy}
\xymatrix@C=10mm@R=5mm{
{\begin{matrix}
H^3\left(\bY,B_{1,0,1} \right)_6
 \\\oplus \\
H^3\left(\bY,B_{2,1,0} \right)_{15}
\end{matrix}} 
   \\ 
 0 & 0  \\
0 &  
H^1\left(\bY,B_{1,1,0} \right)_{2}
& 0  \\
0 & 
{\begin{matrix}
H^0\left(\bY,B_{0,0,1} \right)_{21}
\\\oplus\\ 
H^0\left(\bY,B_{1,1,0} \right)_{6}
\end{matrix}}   \ar[r]^{\bQb_W}  &  
{\begin{matrix}
H^0\left(\bY, B_{0,1,0}\right)_{295}
\end{matrix}}
}
\save="x"!LD+<-6mm,0pt>;"x"!RD+<35pt,0pt>**\dir{-}?>*\dir{>}\restore
\save="x"!LD+<110mm,-3mm>;"x"!LU+<110mm,2mm>**\dir{-}?>*\dir{>}\restore
\save!RD+<-95mm,-4mm>*{-\ff{5}{2}}\restore
\save!RD+<-55mm,-4mm>*{-\ff{3}{2}}\restore
\save!RD+<-16mm,-4mm>*{-\half}\restore
\save!RD+<12mm,-3mm>*{p}\restore
\save!CL+<107mm,34mm>*{U}\restore
\end{xy}
\end{equation}
The bottom row is the only place where we can have cokernel, and for generic superpotential we find $\dim \ker \bQb_W=1$.

\begin{table}[!t]
\renewcommand{\arraystretch}{1.3}
\begin{center}
\begin{tabular}{cc}
\begin{tabular}{|c|c|c|c|c|c|c|c|c|c|}
\hline
$k$ 		&$E_{|k\ra}$	&$\bq_{|k\ra}$		&$\bqb_{|k\ra}$ 		&$\ell_k$	&$\nu_i$			&$\nut_i$		
\\ \hline
$0$		&$0$		&$-\ff{3}{2}$		&$-\ff{3}{2}$			&$0$			&$0$			&$0$							
\\ \hline	
$1$		&$-1$		&$0$			&$-\ff{3}{2}$			&$0$			&$\ff{1}{4}$		&$-\ff{1}{4}$					
\\ \hline
$2$		&$0$		&$\half$			&$-\ff{3}{2}$			&$-4$			&$\ff{1}{2}$		&$-\ff{1}{2}$			
\\ \hline
\end{tabular}&
\begin{tabular}{|c|c|c|c|c|}
\hline
~ 		&$\phi^i$			&$\rho_i$				&$\chi^i$			&$\chib_i$  \\ \hline
$\bq$	&$\ff{1}{2}$		&$-\ff{1}{2}$			&$-\ff{1}{2}$			&$\ff{1}{2}$	\\ \hline
$\bqb$      &$\ff{1}{2}$		&$-\ff{1}{2}$			&$\ff{1}{2}$			&$-\ff{1}{2}$ \\ \hline
\end{tabular}
\end{tabular}
\end{center}
\caption{Quantum numbers for the $X=\cO(-5/2)\oplus\cO(-3/2)\rightarrow \P^3$ model.}
\label{table:orbinums}
\end{table}
Thus, the $E_2$ stage of the spectral sequence is 
\begin{equation}
\begin{matrix}\vspace{10mm}\\E_2^{p,u}:\end{matrix}
\begin{xy}
\xymatrix@C=5mm@R=5mm{
  &
\C^{21}
   \\ 
& 0 & 0 \\
  & 0  & \C^2 & 0  \\
 & 0 & 
\C &  \C^{269} & 0
}
\save="x"!LD+<5mm,0pt>;"x"!RD+<15pt,0pt>**\dir{-}?>*\dir{>}\restore
\save="x"!LD+<43mm,-3mm>;"x"!LU+<43mm,2mm>**\dir{-}?>*\dir{>}\restore
\save!RD+<-41mm,-4mm>*{-\ff{5}{2}}\restore
\save!RD+<-28mm,-4mm>*{-\ff{3}{2}}\restore
\save!RD+<-15mm,-4mm>*{-\ff{1}{2}}\restore
\save!RD+<-2mm,-4mm>*{\ff{1}{2}}\restore
\save!RD+<4mm,-4mm>*{p}\restore
\save!CL+<40mm,20mm>*{U}\restore
\end{xy}
\end{equation}
All higher differentials vanish, and the spectral sequence degenerates at the $E_2$ term. We then count 271 chiral and 21 antichiral states corresponding to massless $\Le_6$ singlets.
We also computed by similar methods the number of charged singlets, $h^{2,1} = 90$ and $h^{1,1}=2$, corresponding to the (2,2) moduli, which we can subtract from the total number of neutral singlets 
to find $\cM=200$.

\subsubsection*{A higher differential ?}
It is worth noting that the spectral sequence for computing the number of $\rep{1}_2\subset\mathbf{27}$ states degenerates only at the $E_4$ term, giving
us an example of a possible higher differential. At zero energy and $\bq=2$ we have
\begin{equation}
\begin{matrix}\vspace{10mm}\\E_1^{p,u}:\end{matrix}
\begin{xy}
\xymatrix@C=5mm@R=6mm{
   0 \\ 
 {H^2(\bY,B^1_{3,0,0})}_{1}  &  0 \\
   0 & 0
  & 0  \\
  0  &  {H^0(\bY,B_{2,0,0})}_{120} 
\ar[r]^-{\bQb_W} &
 {H^0(\bY,B_{1,0,0})}_{905} 
\ar[r]^-{\bQb_W} &
 {H^0(\bY,B_{0,0,0})}_{875} 
}
\save="x"!LD+<-2mm,0pt>;"x"!RD+<15pt,0pt>**\dir{-}?>*\dir{>}\restore
\save="x"!LD+<101mm,-3mm>;"x"!LU+<101mm,2mm>**\dir{-}?>*\dir{>}\restore
\save!RD+<-118mm,-4mm>*{-\ff{5}{2}}\restore
\save!RD+<-85mm,-4mm>*{-\ff{3}{2}}\restore
\save!RD+<-50mm,-4mm>*{-\half}\restore
\save!RD+<-15mm,-4mm>*{\half}\restore
\save!RD+<5mm,-3mm>*{p}\restore
\save!CL+<98mm,22mm>*{U}\restore
\end{xy}
\end{equation}
Trivially $d_2=0$, thus $E_3=E_2$, but there is one more map we have to compute, in fact
\begin{equation}
\begin{matrix}\vspace{10mm}\\E_3^{p,u}:\end{matrix}
\begin{xy}
\xymatrix@C=7mm@R=7mm{
  0 \\ 
 \C \ar[rrrdd]^-{d_3}& 0\\
   0  & 0 &   0  \\
  0 & 0 & 0 & \C^{90}
}
\save="x"!LD+<00mm,0pt>;"x"!RD+<15pt,0pt>**\dir{-}?>*\dir{>}\restore
\save="x"!LD+<31mm,-3mm>;"x"!LU+<31mm,2mm>**\dir{-}?>*\dir{>}\restore
\save!RD+<-41mm,-4mm>*{-\ff{5}{2}}\restore
\save!RD+<-29mm,-4mm>*{-\ff{3}{2}}\restore
\save!RD+<-18mm,-4mm>*{-\ff{1}{2}}\restore
\save!RD+<-5mm,-4mm>*{\ff{1}{2}}\restore
\save!RD+<4mm,-4mm>*{p}\restore
\save!CL+<28mm,22mm>*{U}\restore
\end{xy}
\end{equation}
Let us recall that an element $b\in \cH$ represents a cohomology class in $E_3$ if there exist $c_1,c_2 \in \cH$ such that 
\begin{align}
\bQb_0 b &=0~, & \bQb_W b &= \bQb_0 c_1 ~ , &\bQb_W c_1 &= \bQb_0 c_2~ ,
\end{align}
and $d_3$ on the cohomology class $[b]_3$ is given by 
\begin{align}\label{d3coho}
d_3[b]_3 = [\bQb_W c_2]_3 ~.
\end{align}
Thus, we just chase down the state $\etab^{\Jb}\etab^{\Kb}V^I_{\Jb\Kb}\chib_1\chib_2\chib_I |1\ra  \in E_3^{-5/2,3}$ as prescribed in (\ref{d3coho})
\begin{align}
\xymatrix@C=5mm{0 & 0\\
\etab^{\Jb}\etab^{\Kb}V^I_{\Jb\Kb}\chib_1\chib_2\chib_I |1\ra   \ar[r]^-{\bQb_W}  \ar[u]^-{\bQb_0}&
\etab^{\Jb}\etab^{\Kb}V^I_{\Jb\Kb}\epsilon^{\alpha\beta\gamma} \p_\alpha W \chib_\beta\chib_\gamma |1\ra  \ar[u]^-{\bQb_0}\\
&\etab^{\Jb}S^I_{\Jb}\epsilon^{\alpha\beta\gamma} \p_\alpha W \chib_\beta\chib_\gamma |1\ra
\ar[u]^{\bQb_0} \ar[r]^-{\bQb_W}& 
 \etab^{\Jb}S^I_{\Jb}\epsilon^{\alpha\beta\gamma} \p_\alpha W \p_{[\beta} W\chib_{\gamma]} |1\ra \\
& & R^I \epsilon^{\alpha\beta\gamma} \p_\alpha W \p_{[\beta} W\chib_{\gamma]}  |1\ra
  \ar[u]^{\bQb_0}  \ar[r]^-{\bQb_W}&  
0 ~ .
}
\end{align}
The coefficients satisfy 
\begin{align}
V^I_{\Jb\Kb} &= -(\pb S)^I_{\Jb\Kb}~, &S^I_{\Jb} &= - (\pb R)^I_{\Jb}~.
\end{align}
We just showed that $d_3$, while in principle allowed, vanishes, and the spectral sequence degenerates at the $E_4=E_2$ term. In this sector we
count $h^{2,1}=90$ and $h^{1,1}=1$ and the ``missing" K\"ahler modulus is to be found in the $k=3$ sector, as expected.

\subsection{A positive line bundle}

For our last example we consider $X=\cO(-3,-3)\oplus\cO(1,1)$ and $B=\mathbb{F}_0$. The novelty here is that we allow a positive line bundle over a non-projective base.

A non degenerate superpotential is given by
\begin{align}
W= (\phi^1)^4 S_{[12,12]} + (\phi^1)^3\phi^2 S_{[8,8]} + (\phi^1\phi^2)^2 S_{[4,4]} + \phi^1(\phi^2)^3 S_{[0,0]}~,
\end{align}
where $S_{[m,n]} \in \Gamma(\mathbb{F}_0,\cO(m,n))$ 
and the quantum numbers for this theory are listed in table \ref{table:F0nums}.
Studying the (R,R) sectors we find $h^{1,1}=3$ and $h^{2,1}=243$, and to count the remaining $\Le_6$ singlets we need to consider the $k=1$ and $k=3$ sectors.

\subsubsection*{$k=1$ sector}
In the first twisted sector the spectral sequence at $\bq=0$ is
\begin{equation}
\begin{matrix}\vspace{20mm}\\E_1^{p,u}:\end{matrix}
\begin{xy}
\xymatrix@C=15mm@R=5mm{
{\begin{matrix}
  {H^2(\bY,B_{1,1,0})}_{39} \\\oplus\\
  {H^2(\bY,B_{0,0,1})}_{9} 
\end{matrix}}
 \ar[r]^-{\bQb_W} &
  {H^2(\bY,B_{0,1,0})}_{39} \\
  {H^1(\bY,B_{1,1,0})}_{10}
 \ar[r]^-{\bQb_W} &
  {H^1(\bY,B_{0,1,0})}_2 \\
 {\begin{matrix}
  {H^0(\bY,B_{1,1,0})}_{63} \\\oplus\\
    {H^0(\bY,B_{0,0,1})}_{27}
  \end{matrix}}
\ar[r]^-{\bQb_W}
  &
  {H^0(\bY,B_{0,1,0})}_{825} 
}
\save="x"!LD+<-2mm,0pt>;"x"!RD+<50pt,0pt>**\dir{-}?>*\dir{>}\restore
\save="x"!LD+<75mm,-3mm>;"x"!LU+<75mm,2mm>**\dir{-}?>*\dir{>}\restore
\save!RD+<15mm,-3mm>*{p}\restore
\save!CL+<72mm,28mm>*{U}\restore
\save!RD+<-59mm,-4mm>*{-\ff{3}{2}}\restore
\save!RD+<-18mm,-4mm>*{-\half}\restore
\end{xy}
\end{equation}
It is not hard to verify that both the maps $\bQb_W\big|_{U=1}$ and $\bQb_W\big|_{U=2}$ are surjective for sufficiently generic $W$, and as we already saw in the discussion about the general $k=1$ sector, there is only one state at $\bqb=-\ff{3}{2}, u=0$ in the kernel of $\bQb_W$.
The spectral sequence degenerates at the $E_2$ term
\begin{equation}
\begin{matrix}\vspace{10mm}\\E_2^{p,u}:\end{matrix}
\begin{xy}
\xymatrix@C=10mm@R=5mm{
 \mathbb{C}^9  \\
 \mathbb{C}^8 & 0   \\
 \mathbb{C} & \mathbb{C}^{736} &
}
\save="x"!LD+<-6mm,0pt>;"x"!RD+<0pt,0pt>**\dir{-}?>*\dir{>}\restore
\save="x"!LD+<30mm,-3mm>;"x"!LU+<30mm,2mm>**\dir{-}?>*\dir{>}\restore
\save!RD+<-1mm,-4mm>*{p}\restore
\save!CL+<27mm,15mm>*{U}\restore
\save!RD+<-37mm,-4mm>*{-\ff{3}{2}}\restore
\save!RD+<-19mm,-4mm>*{-\half}\restore
\end{xy}
\end{equation}
Thus we count $744$ chiral and $9$ anti-chiral massless $\Le_6$-singlets and one vector.

\begin{table}[!t]
\renewcommand{\arraystretch}{1.3}
\begin{center}
\begin{tabular}{cc}
\begin{tabular}{|c|c|c|c|c|c|c|c|c|c|}
\hline
$k$ 		&$E_{|k\ra}$		&$\bq_{|k\ra}$		&$\bqb_{|k\ra}$ 		&$\ell_k$			&$\nu_i$			&$\nut_i$			
\\ \hline
$0$		&$0$		&$-\ff{3}{2}$	&$-\ff{3}{2}$	&$0$			&$0$			&$0$			
\\ \hline
$1$		&$-1$		&$0$		&$-\ff{3}{2}$	&$0$			&$\ff{1}{8}$		&$-\ff{3}{8}$				
\\ \hline
$2$		&$0$		&$\half$		&$-\ff{3}{2}$	&$(-2,-2)$			&$\ff{1}{4}$		&$-\ff{3}{4}$		
\\ \hline
$3$		&$-\ff{3}{4}$	&$-\half$		&$-1$		&$0$			&$\ff{3}{8}$		&$-\ff{1}{8}$	
\\ \hline
$4$		&$0$		&$-1$		&$-1$		&$(-2,-2)$			&$\ff{1}{2}$		&$-\ff{1}{2}$		
\\ \hline
\end{tabular}&
\begin{tabular}{|c|c|c|c|c|}
\hline
~ 		&$\phi^i$			&$\rho_i$				&$\chi^i$			&$\chib_i$  \\ \hline
$\bq$	&$\ff{1}{4}$		&$-\ff{1}{4}$			&$-\ff{3}{4}$			&$\ff{3}{4}$	\\ \hline
$\bqb$      &$\ff{1}{4}$		&$-\ff{1}{4}$			&$\ff{1}{4}$			&$-\ff{1}{4}$ \\ \hline
\end{tabular}
\end{tabular}
\end{center}
\caption{Quantum numbers for the $X=\cO(-3,-3)\oplus\cO(1,1)\rightarrow \mathbb{F}_0$ model.}
\label{table:F0nums}
\end{table}

\subsubsection*{$k=3$ sector}
In the $k=3$ sector all the fields are ``light'', $L_{|3\ra}$ is trivial, and the geometry is again encoded in the full $\bY_{\!\!3} = \bY$. The spectral sequence starts then as 
 \begin{equation}
\begin{matrix}\vspace{15mm}\\E_1^{p,u}:\end{matrix}
\begin{xy}
\xymatrix@C=15mm@R=5mm{
{\begin{matrix} 
{H^2(\bY,\Sym^2 T_{\bY})}_{27} \\\oplus\\
 {H^2(\bY,\wedge^2 T_{\bY}\otimes T^\ast_{\bY})}_6 
 \end{matrix}}  \ar[r]^-{\bQb_W}
 &  {H^0(\bY,\cO)}_1 \\
0 & 0   \\
  {H^0(\bY,\Sym^2 T_{\bY})}_9  
 \ar[r]^-{\bQb_W}
 & 
 {H^0(\bY,\cO)}_{58}
}
\save="x"!LD+<-2mm,0pt>;"x"!RD+<35pt,0pt>**\dir{-}?>*\dir{>}\restore
\save="x"!LD+<76mm,-3mm>;"x"!LU+<76mm,2mm>**\dir{-}?>*\dir{>}\restore
\save!RD+<-56mm,-4mm>*{-\ff{3}{2}}\restore
\save!RD+<-12mm,-4mm>*{-\ff{1}{2}}\restore
\save!RD+<11mm,-4mm>*{p}\restore
\save!CL+<73mm,21mm>*{U}\restore
\end{xy}
\end{equation}
It can be shown that the map $\bQb_W\big|_{U=0}$ is injective while the map $\bQb_W\big|_{U=2}$ is surjective. Therefore
 the second stage of the spectral sequence is
\begin{equation}
\begin{matrix}\vspace{10mm}\\E_2^{p,u}:\end{matrix}
\begin{xy}
\xymatrix@C=10mm@R=5mm{
\mathbb{C}^{32} & 0   \\
0 & 0  \\
 0& \mathbb{C}^{49} &
}
\save="x"!LD+<-2mm,0pt>;"x"!RD+<0pt,0pt>**\dir{-}?>*\dir{>}\restore
\save="x"!LD+<30mm,-3mm>;"x"!LU+<30mm,2mm>**\dir{-}?>*\dir{>}\restore
\save!RD+<-1mm,-4mm>*{p}\restore
\save!CL+<27mm,15mm>*{U}\restore
\save!RD+<-36mm,-4mm>*{-\ff{3}{2}}\restore
\save!RD+<-17mm,-4mm>*{-\half}\restore
\end{xy}
\end{equation}
The spectral sequence degenerates at the $E_2$ term, and we find 49 chiral and 32 anti-chiral singlets.

Summarizing, we count 834 massless chiral $\Le_6$-singlets and once we subtract the moduli we obtain $\cM=588$.


\section{Discussion} \label{s:discuss}

We have described a class of perturbative vacua for heterotic string compactifications and a limit in which their properties are computable.  We have illustrated these computations in models with (2,2) world-sheet supersymmetry, although the methods clearly extend to more general (0,2) theories.  

Our class of (2,2) models fits in with a number of other constructions.  To describe this
we proceed in increasing dimension $d$ of the base $B$ and assume this is Fano.  For $d=1$ this means $B=\P^1$ and the $c=6$ LGO theory on the fiber determines a one-parameter family of K3 compactifications.   Models with no large radius limit in the K\"ahler moduli space, such as the first example in section~\ref{s:Examples}, are obtained when the monodromies of the family are not simultaneously geometrical in any duality frame.  It seems likely that any such model would be obtained as a limit in some GLSM, but we have not shown this.

For $d=2$ the base is a del Pezzo surface and the $c=3$ LGO theory on the fiber can be interpreted as determining in Weierstrass form an elliptic fibration over $B$.   This can be smooth if the discriminant is nonsingular in $B$, in which case the model will have a large-radius limit.   It is not clear how to construct a GLSM embedding for a hybrid with a non-toric base.

For $d=3$ there are many possible choices for $B$, but the $c=0$ LGO theory is quadratic, and hence appears to be trivial.  Since the fiber fields are massive at generic points on the base, one might think the low-energy theory would be a NLSM with target space $B$, but this cannot be correct, as this would not be conformally invariant.  This na\"ive discussion omits the orbifold action.  Since at low energy there are no excitations in the fiber direction, one can try~\cite{Addington:2012zv} to describe the resulting model as a NLSM with target space a double cover of $B$ branched over the singular locus of $W(y,\phi)$ {\it considered as a function of $\phi$ only\/}.   This leads to a geometric interpretation of the limiting point we called the hybrid limit.  It is not directly related to a symplectic quotient construction and, if the model has a large-radius limit, it is not birational to the target space at this limit.   The relationship between the two descriptions is unclear.
It would be interesting to study, among other things, the behavior of the D-brane spectrum and moduli in a type-IIA compactification near such a hybrid limit.  

The models we have studied have been ``good'' hybrids, in which the R-symmetry does not act on the base.   Limiting points of GLSMs often produce hybrids for which this does not hold.   The hybrid limit for ``good'' hybrids is expected to lie at infinite distance in the moduli space of SCFTs; it should be possible to determine the approximate moduli space metric in the hybrid limit.  We expect that the approximation should improve as the hybrid limit is approached and the distance to the hybrid limit deep in the K\"ahler cone of $B$ will diverge.  It would be interesting to verify this in detail.  In~\cite{Aspinwall:2009qy} ``pseudo'' hybrids were defined as hybrid limits lying at finite distance; the behavior of the D-brane spectrum near these limits was found to be quite different from that expected near a ``good'' hybrid.    It seems natural to conjecture that ``good'' hybrids and ``true'' (not ``pseudo'') hybrid limits coincide.

Although we focused on models with (2,2) world-sheet supersymmetry, the methods extend naturally to a much larger class of models with (0,2) supersymmetry.   This larger class presents an array of interesting questions.   As a first foray in this direction, the massless $\Le_6$ singlets in (NS,R) 
sectors belong to (anti-) chiral multiplets containing massless scalars.  Expectation values for these represent marginal deformations of the world-sheet SCFT preserving (0,2) supersymmetry.   We do not at present have effective techniques to determine which of these are exactly marginal, and the structure of the moduli space of (0,2) SCFTs is still largely unknown.    

In general one expects~\cite{Dine:1986zy,Distler:1987ee}  that away from the hybrid limit the (0,2) models we construct will be destabilized by world-sheet instantons wrapping cycles in $B$.  In some classes of models this expectation has been thwarted, and the anticipated corrections are absent~\cite{Silverstein:1995re,Basu:2003bq,Beasley:2003fx}.  Even in cases in which no known argument precludes such corrections they have been found less generally than one might expect~\cite{Aspinwall:2010ve,Aspinwall:2011us}.   It would be very interesting to investigate this issue in the context of hybrid models, in which the structure of the relevant instantons --  associated to rational curves in $B$ rather than in a Calabi--Yau threefold, may provide a simpler context for their study.

More generally, we can construct (0,2) hybrid models that are not deformations of (2,2) models by taking the left-moving fermions to be sections of a holomorphic bundle $\cE\to \bY$ and a (0,2) superpotential given by a section  $J \in \Gamma(\cE^\ast)$ with $J^{-1}(0) = B$.   It is to be expected that most such models will not have a limit in which they are described by a (0,2) NLSM or one in which they reduce to a (0,2) LGO theory, so that these will determine a large class of new perturbative vacua of the heterotic string.   These models will be considered in a forthcoming work.

\appendix

\section{Hybrid geometry : an example} \label{app:simplehybrid}
Let $B = \P^1$ and take $\bY$ to be the total space of $X =  \cO(-2) \to \P^1$.  We cover $\bY$ by two patches ${\cU}_{u}$ and $\cU_{v}$, with local coordinates $(u,\phi_u)$ and $(v,\phi_v)$, respectively :
\begin{align}
u = v^{-1}, \qquad \phi_u = v^2\phi_v \qquad\text{on} \quad {\cU}_{u}\cap{\cU}_v = \C^\ast.
\end{align}
The projection $\pi :\bY \to B$ is simply $(u,\phi_u) \to u$ and $(v,\phi_v) \to v$ in the two patches. 
The transition function for  $\sigma = \sigma^1_u \p_{u} + \sigma^2_u \p_{{\phi_u}}$, a section of $T_{\bY}$, is 
\begin{align}
\begin{pmatrix}
\sigma^1_u & \sigma^2_{u} 
\end{pmatrix} = 
\begin{pmatrix}
\sigma^1_v & \sigma^2_{v} 
\end{pmatrix}
\begin{pmatrix}
-v^{-2} & 2v\phi_v \\
0 & v^2 
\end{pmatrix}
~.
\end{align}
$T_{\bY}$ belongs to a family of rank $2$ holomorphic bundles $\cV_\ep\to {\bY}$ with transition function
\begin{align}
\label{eq:extrans}
\begin{pmatrix}
\sigma^1_u & \sigma^2_{u} 
\end{pmatrix} = 
\begin{pmatrix}
\sigma^1_v & \sigma^2_{v} 
\end{pmatrix} M_{\ep}~, \qquad
M_{\ep} \equiv
\begin{pmatrix}
-v^{-2} & 2\ep v\phi_v \\
0 & v^2 
\end{pmatrix}
~.
\end{align}
When $\ep = 0$ the bundle splits: $\cV_{\ep=0} = \pi^\ast \cO(2)\oplus\pi^\ast \cO(-2)$; more generally $\cV_{\ep}$ is an irreducible rank $2$ bundle over $\bY$. 

An example of a quasi-homogeneous superpotential depending on a parameter $\alpha$ is
\begin{align}
W_u = (\alpha+u^8) \phi_u^4~,\qquad W_v = (\alpha v^8+1) \phi_v^4~.
\end{align}
Clearly $W_u = W_v$ on the overlap.  Computing the gradient in the two patches, we obtain
\begin{align}
dW_u = 8 u^7 \phi_u^4 du + 4 (\alpha+u^8) \phi_u^3 d\phi_u~,\qquad
dW_v =8 \alpha v^7 \phi_v^4 dv + 4 (\alpha v^8+1) \phi_v^3 d\phi_v~.
\end{align}
It is then easy to see that for $\alpha \neq 0$ we have $dW^{-1}(0) = B$.  A more general superpotential respecting the same quasi-homogeneity is
\begin{align}
W_u = S_u(u) \phi_u^4~,\qquad W_v = S_v(v) \phi_v^4~,
\end{align}
where $S_{u,v}$ is the restriction of $\Sigma \in H^0(B,\cO(8))$ to $\cU_{u,v}$.  The potential condition is satisfied for generic choices of $\Sigma$.

We can see how the fibration affects the naive chiral ring $R_p$ of the LG fiber theory over a point $p\in B$: $\dim R_p$ jumps in complex co-dimension $1$ but stays finite if the potential condition is satisfied.  In our example $R_u = \{1,\phi_u,\phi_u^2\}$ for $u^8+\alpha \neq 0$, while at the $8$ special points $R = \{1,\phi_u,\phi_u^2,\phi_u^3\}$.  If $\alpha = 0$ then the potential condition is violated, and $\dim R_0 = \infty$. 

\subsubsection*{A (0,2) deformation}
Taking the left-moving bundle to be $\cE = \cV_{\ep}$, we obtain a class of (0,2) theories.  The most general (0,2) superpotential that respects the same quasi-homogeneity as $dW$, $J \in \Gamma(\cE^\ast)$, takes the form
\begin{align}
J_u =
\begin{pmatrix}
T_u(u)\phi_u^4 \\ 4S_u(u) \phi_u^3
\end{pmatrix}, \qquad
J_v =\begin{pmatrix}
T_v(v)\phi_v^4 \\ 4S_v(v) \phi_v^3
\end{pmatrix}~, \qquad
\end{align}
where $S$ and $T$ are holomorphic functions constrained by $J_u = M_\ep^{-1} J_v$ when $u\neq v$.  $S_{u,v}$ are restrictions of $\Sigma$ as above, while $T_{u,v}$ are given by
\begin{align}
T_u(u) = -\left.\Sigmat\right|_{u} + 8\ep u^{-1} \left(S_u(u) - S_u(0) \right)~, \qquad
T_v(v) = \left.\Sigmat\right|_{v} + 8\ep v^7 S_u(0)~,
\end{align}
where $\Sigmat\in H^0(B,\cO(6))$.
The potential condition is satisfied for generic $\Sigma$ and $\Sigmat$.  Setting $\ep=1$ and $T_u = \p_u S_u$, we recover the (2,2) potential from above.  On the other hand, taking $\ep =0$, we see that $T$ is just given by restriction of holomorphic sections of $\cO(6)$.

We can compare the number of holomorphic deformation parameters in the (2,2) or (0,2) superpotentials.  $W$ depends on $9$ holomorphic parameters specifying  section $\Sigma$.  The more generic (0,2) superpotential $J$, on the other hand, depends on $16$ parameters, independent of $\ep$; as a check, we see that demanding that $J$ is integrable to $W$ reduces the parameters to $9$.

\subsubsection*{Metrics for $\bY$ and $\cE$}
It is well known that $\bY$ admits an ALE K\"ahler Ricci-flat metric with K\"ahler potential~\footnote{Constructions of such metrics for line bundles over $\P^{n-1}$, which generalize the classic work of Eguchi and Hanson~\cite{Eguchi:1980jx}, go back to~\cite{Calabi:1979rf,Freedman:1981rf}; ~\cite{Higashijima:2002jt} gives an elegant generalization for line bundles over symmetric spaces.  These are also the only explicitly known ALE metrics with $\SU(n)$ $n\ge 3$ holonomy~\cite{Joyce:2000cm}.}
\begin{align}
K_{\text{CY}} =  \sqrt{1+L} + \frac{1}{2} \log\frac{ \sqrt{1+L}-1}{\sqrt{1+L}+1}~, \qquad
L \equiv 4  \phi\phib(1+u\ub)^2~.
\end{align}
This is obviously well-defined with respect to the patching.  To leading order in the fiber coordinates, we find that up to irrelevant constants
\begin{align}
K_{\text{CY}} = K + O(|\phi|^4), \qquad K =  \log(1+u\ub) + (1+u\ub)^2 \phi\phib~.
\end{align}
$K$ leads to a complete non-Ricci-flat metric on $X$:
\begin{align}
g_X =\begin{pmatrix} g_{u\ub} & g_{u\phib} \\ g_{\phi\ub} & g_{\phi\phib} \end{pmatrix} 
= \begin{pmatrix}   (1+u\ub)^{-2} + 2 (1+2u\ub)\phi\phib & 2\ub\phi(1+u\ub)\\
2 u\phib(1+u\ub) &  (1+u\ub)^2 \end{pmatrix}.
\end{align}
To $O(|\phi|^4)$ this agrees with the K\"ahler metric obtained by symplectic reduction from $\C^3$.

We can also endow $\cE$ with a Hermitian metric.  In our example with $\cE = \cV_\ep$, a convenient choice is
\begin{align}
(\sigma,\tau) \equiv \sigma \cG \overline{\tau} ,\qquad
\cG = \begin{pmatrix}  (1+u\ub)^{-2} + 2\ep\epb (1+2u\ub)\phi\phib & 2\ep\ub\phi(1+u\ub)\\
2\epb u\phib(1+u\ub) & (1+u\ub)^2 \end{pmatrix}.
\end{align}
Setting $\ep =1$, we obtain a Hermitian, in fact K\"ahler,  metric on $T_{\bY}$.  Setting $\phi =0$ we obtain  the bundles restricted to $B$.  As we might expect, $T_{\bY}|_{B} = \cV_{\ep}|_{B} = \cO(2) \oplus \cO(-2)$.

The explicit Ricci-flat metric on $\bY$ is fairly complicated, and generalizations to other spaces are typically not available.  Fortunately, we do not need the explicit form of the metric for our analysis: by construction the superpotential restricts low energy field configurations to $B$, and the details of the metric on $\bY$ away from the base become irrelevant to the IR physics.

\section{Vertical Killing vectors} \label{app:VKill}
In this appendix we examine holomorphic vertical Killing vectors on $\bY$ and prove that with our assumptions they act homogeneously on the fiber directions.

Let $V= V^\alpha \pp{y^\alpha} + \text{c.c.}$ be an holomorphic vector field on $\bY$, i.e. $V^\alpha_{,\betab} = 0$.  Then the Killing equation for a K\"ahler metric $g_{\alpha\betab}$ takes the form
\begin{align}
\p_\gamma (g_{\alpha\betab} V^\alpha) + \p_{\betab} (g_{\gamma\alphab}\Vb^{\alphab}) = 0~.
\end{align}
Using the base/fiber decomposition $y^\alpha = (u^I,\phi^i)$, the hybrid metric has components
\begin{align}
g_{I\Jb} = G_{I\Jb} + \phi h_{I\Jb}\phib,\qquad
g_{i\Jb} = h_{i\mb \Jb} \phib^{\mb},\qquad
g_{I\jb} = \phi^m h_{m\jb I},\qquad
g_{i\jb} = h_{i\jb}~.
\end{align}
Since $V$ is vertical, we have $V = V^i \pp{\phi^i} + \text{c.c.}$, and a moment's thought shows that $V^i(u,\phi)$ transforms as a section of $\pi^\ast(X)$.  In this case the Killing equation reduces to
\begin{align}
\p_\gamma( g_{i\betab} V^i) + \p_{\betab} (g_{\gamma\ib} \Vb^{\ib} ) = 0~,
\end{align}
and decomposing it further along base/fiber directions leads to two non-trivial conditions.  First, from $\betab,\gamma = \jb,k$ we obtain
\begin{align}
\p_k V^i + h^{\jb i} \p_{\jb} \Vb^{\ib}_{\jb} h_{k\ib} = 0~.
\end{align}
Since $h$ is $\phi$-independent and  $\p_m \Vb^{\ib}_{\jb} = 0$, we conclude that
\begin{align}
V^i = A^i_k(u) \phi^k + B^i(u),\qquad \Ab^{\ib}_{\kb} = (A^i_k)^\ast = -h^{\ib i} A^{k}_{i} h_{k\kb}~.
\end{align}
The latter restriction on $A \in H^0(B, X\otimes X^\ast)$, combined with its holomorphy leads to  $D_J A = 0$.
The remaining non-trivial conditions are obtained by taking $\betab,\gamma = J, \kb$ in the Killing equation, and they lead to $D_J B = 0$ for $B \in H^0(B, X)$.

So, we have learned that vertical automorphic Killing vectors are characterized by covariantly constant sections $A \in \Gamma(X\otimes X^\ast)$ and $B \in \Gamma (X)$, with the additional restriction
\begin{align}
(A^i_k)^\ast = -h^{\ib i} A^{k}_{i} h_{k\kb}~.
\end{align}
In fact, we can always shift away the global section $B$ by a redefinition of the $\phi^i$; moreover, for a generic choice of metric $h$ the only solution for $A$ is a diagonal anti-Hermitian $u$-independent matrix; demanding $\cL_V W = W$ will fix the eigenvalues (up to an overall $i$) to be the charges $q_i$.

\section{A little sheaf cohomology} \label{app:sheaf}
In this section we present some useful results for reducing sheaf cohomology on $\bY$ to computations on the base $B$ in the case that $X =\oplus_i L_i$.  In order to compute $\bQb_0$ cohomology we need an effective method to evaluate
\begin{align}
\label{eq:sheafwant}
 H^\bullet_{\br}(\bY, \pi^\ast(\cE)\otimes \wedge^s T_{\bY} \otimes \wedge^t T^\ast_{\bY}),
\end{align}
where $\cE$ is some bundle (or more generally sheaf) on $B$, and $\br$ is the restriction to fine grade $\br$.
Recall that the grading $\br\in \Z^{n}$ assigns to every monomial $\prod_i\phi_i^{r_i}$ grade $\br = (r_1,\ldots,r_n)$; in particular $\phi_i$ has grade $\bx_i$ with $(\bx_i)_j = \delta_{ij}$.
Since $\bY$ is non-compact the grade restriction is necessary to obtain a well-posed counting problem.  For instance, the structure sheaf $\cO_{\bY}$ clearly has infinite-dimensional cohomology group $H^0(\bY,\cO_{\bY})$.  

\subsubsection*{Graded cohomology of a pulled-back sheaf}
Suppose $s=t=0$ in~(\ref{eq:sheafwant}).  As we now show,
\begin{align}
\label{eq:sheafpullback}
H^\bullet_{\br}(\bY,\pi^\ast(\cE)) \simeq H^\bullet(B, \cE\otimes \LL_{\br}),
\end{align}
where $\LL_{\br} \to B$ is the line bundle $\LL_{\br} \equiv \otimes_i (L_i^\ast)^{r_i}$.

The proof follows from the basic geometry.  First, to describe the line bundles $L_i\to B$, we work with a cover $\cU = \{ U_a\}$ for $B$ with local coordinates $u_a^I$ in each patch, so that on overlaps $U_{ab}\neq \emptyset$ sections of $L_i$ satisfy 
\begin{align}
\lambda^i_b (u_b) = \lambda^i_a(u_a) g^i_{ab}(u_a)~,
\end{align}
where the $g^i_{ab}$ are the transition functions defining the bundle $L_i$.
The sections $\sigma_a$ of a sheaf $\cE \to B$ satisfy
\begin{align}
\sigma_b (u_b) = \sigma_a(u_a) G_{ab}(u_a)~,
\end{align}
where the $G_{ab}$ are the transition functions for $\cE$, and sections of 
$\pi^\ast(\cE) \to \bY$ patch as
\begin{align}
\label{eq:overY}
\sigma_b (u_b,\phi_b) = \sigma_a(u_a,\phi_a) G_{ab}(u_a)~,
\end{align}
with $\phi^i_b = \phi^i_a g^i_{ab}(u_a)~$.  
Since the transition functions for $\pi^\ast(\cE)$ are identical to the transition functions for $\cE$ over $B$, at fixed grade~(\ref{eq:overY}) takes the form
\begin{align}
\prod_i (\phi_b^i)^{r_i} \xi_b(u_b) =
\prod_i (\phi_a^i)^{r_i} \xi_a(u_a) G_{ab}(u_a) \iff
\xi_b(u_b) = \xi_a(u_a) G_{ab}(u_a) \prod_i \left[ g^i_{ab}(u_a)\right]^{-r_i}~.
\end{align}
Hence the space of sections of $\pi^\ast(\cE)_{\br}$ over $\bY$ is isomorphic to the space of sections of $\cE\otimes \LL_{\br}$ over $B$.  The grading is compatible with \v{C}ech cohomology (i.e. with defining chains for higher intersections $U_{a_1\cdots a_k}$ and taking cohomology of the \v{Cech} differential), and~(\ref{eq:sheafpullback}) holds.

\subsubsection*{The tangent bundle}
Having reduced the graded cohomology of a pull-back sheaf to a cohomology problem on the base, we now turn to the tangent bundle.  This is of course not in general the pull-back of a sheaf from $B$, as we explictly saw in appendix~\ref{app:simplehybrid}.\,  However, $T_{\bY}$ fits into a short exact sequence
\begin{align}
\xymatrix{0 \ar[r]& \pi^\ast(X) \ar[r] & T_{\bY} \ar[r] &\pi^\ast(T_B) \ar[r] & 0}~.
\end{align}
This is easy to see explicitly.  In an open neighborhood $U_a$ a vector field $\Sigma$ takes the form
\begin{align}
\Sigma_a = V_{a} \pp{u_a} + \nu_a \pp{\phi_a},
\end{align}
and on overlaps $U_{ab}$ 
\begin{align}
V_b = V_a \ff{\p u_b}{\p u_a},\qquad \nu_b = \nu_a g_{ab} + \phi_a \cL_V g_{ab}~,
\end{align} 
where $g_{ab}$ are the transition functions for $X$.  Hence, we see that a section $\nu$ of $X$ lifts to a section of $T_{\bY}$ with $V=0$, while a section of $T_{\bY}$ at $\phi = 0$ yields a section of $T_B$.

This short exact sequence can be decomposed with respect to the fine grading.  Working again in the case $X = \oplus_i L_i$, the transition functions for sections of $T_{\bY}$ can be written explicitly as 
\begin{align}
(\sigma^0_b,\sigma^1_b,\ldots,\sigma^n_b) =
(\sigma^0_a,\sigma^1_a,\ldots,\sigma^n_a) 
\begin{pmatrix}
\frac{\p u_b}{\p u_a} & \phi^1_a \p g^1_{ab} & \phi^2_a \p g^2_{ab} & \cdots & \phi^n_a \p g^n_{ab}\\
0 & g^1_{ab} & 0 & \cdots & 0 \\
0 & 0 & g^2_{ab} & \cdots & 0 \\
\vdots & \vdots & \vdots & \ddots & \vdots \\
0 & 0 & 0 & \cdots & g_{ab}^n
\end{pmatrix}~.
\end{align}
Hence, sections of $T_{\bY}$ also admit a fine grading, which we define
\begin{align}
\label{eq:Tsections}
(\Sigma)_{\br} \equiv (\sigma^0_{\br},\sigma^1_{\br+\bx_1},\sigma^2_{\br+\bx_2},\ldots,\sigma^n_{\br+\bx_n})~.
\end{align}
This means the short exact sequence for $T_{\bY}$ can be decomposed according to $\br$ as
\begin{align}
\label{eq:TSES}
\xymatrix{0 \ar[r]& \oplus_i (\pi^\ast L_i)_{\br+\bx_i} \ar[r] & (T_{\bY})_{\br} \ar[r] &(\pi^\ast T_B)_{\br} \ar[r] & 0}~.
\end{align}
Using the induced long exact sequence on cohomology, together with~(\ref{eq:sheafpullback}), we can evaluate $H^{\bullet}_{\br}(\bY,T_{\bY})$.  Taking appropriate products one can generalize this result to compute all desired cohomology groups in~(\ref{eq:sheafwant}).

We should mention one small subtlety in grading the sections of $T_{\bY}$:  from~(\ref{eq:Tsections}) we see that there can be non-trivial contributions for $r_i =-1$.  More precisely, $(T_{\bY})_{\br} = 0$ whenever any $r_i <-1$ or $r_i=r_j=-1$, and if a single $r_i =-1$ we have
\begin{align}
(T_{\bY})_{\br} = (\pi^\ast L_i)_{\br+\bx_i}~,
\end{align}
in which case $H^{\bullet}_{\br} (\bY,T_{\bY}) = H^{\bullet} (B, \LL_{\br})$.

\subsubsection*{Application to $X = \cO(-2)$ and $B= \P^1$}
In this case the grading is one-dimensional $\br = (r)$, the grading bundle is $\LL_s = (\cO(-2)^\ast)^{s} = \cO(2s)$, and for any $r\ge 0$ the structure sheaf cohomology is 
\begin{align}
H^0_r (\bY,\cO_{\bY}) = H^0(B,\cO(2r)) \simeq \C^{2r+1}~,\quad
H^q_r(\bY,\cO_{\bY}) = 0,\qquad\text{for}~q>0~.
\end{align}
For the tangent sheaf the short exact sequence
\begin{align}
\xymatrix{0 \ar[r]& (\pi^\ast \cO(-2))_{r+1} \ar[r] & (T_{\bY})_{r} \ar[r] &(\pi^\ast \cO(2) )_{r} \ar[r] & 0}
\end{align}
leads to the long exact sequence in cohomology
\begin{align}
\xy {\ar(0.05,-10)*{};(0.05,-11)*{}}; 
\xymatrix{ 
0 \ar[r] & H^0(B,\cO(2r)) \ar[r] & H^0_{r} (\bY,T_{\bY}) \ar[r] & H^0(B,\cO(2r+2)) 
\ar@{-} `d[l]`[llld]  \\
H^1(B,\cO(2r)) \ar[r] & H^1_{r} (\bY,T_{\bY}) \ar[r] & H^1(B,\cO(2r+2)) \ar[r] & 0 
}
\endxy
\end{align}
At grade $0$ we obtain
\begin{align}
\xymatrix{ 
0 \ar[r] & \C \ar[r] & H^0_{0} (\bY,T_{\bY}) \ar[r] & \C^3 
\\
0 \ar[r] & H^1_{0} (\bY,T_{\bY}) \ar[r] & 0 \ar[r] & 0 
}
\end{align}
Hence, $H^0_{0} (\bY, T_{\bY}) \simeq \C^4$, and $H^1_0(\bY,T_{\bY}) = 0$.  More generally, for any non-negative grade
\begin{align}
\label{eq:octictangent}
H^0_r (\bY,T_{\bY}) = H^0_r(B,\cO(2r))\oplus H^0_r(B,\cO(2r+2)) \simeq \C^{4r+4},\qquad
H^1_r(\bY,T_{\bY}) = 0~.
\end{align}

\subsubsection*{A note on horizontal representatives}
In order to evaluate $\bQb_0$ cohomology we needed to study the finely graded Dolbeault cohomology of horizontal forms on $\bY$ valued in a holomorphic sheaf $\cF$.  One might wonder what is the relationship between these horizontal forms and more general Dolbeault classes in  $H^{(0,u)}_{\pb} (\bY,\cF)$.  In fact, every such class has a horizontal representative, which is why our results on finely graded cohomology describe horizontal Dolbeault cohomolgy as well.  This is rather intuitive, since the fiber space is simply $\C^n$ (or $\C^n/\Gamma$ for orbi-bundles), but for completeness we give a sketch of the proof.\footnote{This essentially follows the standard proof~\cite{Bott:1982df} that $H^{k}_{dR}(\R^n,\R) =0$ for $k>0$.}

The statement is trivial at $u=0$, so we consider $u=1$.  Let $\tau \in \ker \pb \cap \Omega^{(0,1)}(\bY,\cO_{\bY})$. In any patch $U_a$ we have
\begin{align}
\tau_a = \omega_{a\Ib} d\ub^{\Ib}_{a} + \sigma_{a\ib} d\phib^{\ib}_{a}~.
\end{align}
We define $\eta_a(u_a,\ub_{a},\phi_{a},\phib_\alpha)$ via the line integral
\begin{align}
\eta_\alpha = \int_0^{\phib_a} ~d\zb^{\ib} \sigma_{a\ib}(u_a,\ub_a,\phi_a,\zb)~.
\end{align}
Since $\pb \tau = 0$ implies $\sigma_{a \ib,\jb} =\sigma_{a\jb,\ib}$, the line integral does not depend on the choice of contour from $0$ to $\phib$; moreover, a change of variables $\zb^{\ib} = \gb^{\ib}_{ba} \yb^{\ib}$ in the integral shows that $\eta_a = \eta_b$ on any $U_{ab}\neq \emptyset$, so that $\eta$ patches to a function on $\bY$. Therefore $\tau' = \tau-\pb \eta$ is a (0,1) horizontal form, and a moment's thought shows that $\pb \tau'=0$ implies that it has a holomorphic dependence on the fiber coordinates.

One can generalize the argument to $u>1$ and more general holomorphic sheaf $\cF \to \bY$.  The analogous construction yields $\eta$, a section of $\Omega^{(0,u-1)}(\bY,\cF)$, such that $\tau'= \tau-\pb \eta$ is a horizontal representative of $[\tau] \in H^{(0,u)}_{\pb}(\bY,\cF)$.

\section{Massless spectrum of a (0,2) CY NLSM} \label{app:02NLSM}
In this appendix we apply the first-order techniques developed in section~\ref{ss:bcbg} to marginal deformations of (0,2) NLSMs with CY target space $B$ and a left-moving $\SU(n)$ bundle $\cV$.  We assume $\ch_2(\cV) = \ch_2(T_B)$ and $\cV$ is a stable bundle.  This ensures that the NLSM is conformally invariant to all orders in $\alpha'$ perturbation theory~\cite{Witten:1985bz,Dine:1986zy}.  Our techniques allow us to determine the massless spectrum to all orders in $\alpha'$.  The results for the (R,R) sector and for the gauge-charged matter are exactly the same as those obtained by a Born-Oppenheimer approach in~\cite{Distler:1987ee}.  However, the massless gauge-neutral chiral matter has to our best knowledge not been studied directly in the NLSM.  The first-order formulation of $\bQb_0$ cohomology turns out to be perfectly suited to this task and should be thought of as a first step in systematically including any non-perturbative world-sheet effects.

In parallel with the analysis of the $k=1$ sector in section~\ref{ss:k1}, we first list the operators that can give rise to massless singlets.  We need to slightly alter our notation in comparison to the $T_B = \cV$ analysis of section \ref{ss:k1};   just in this appendix we use $I,J,\ldots$ for tangent/cotangent indices, while the $\alpha,\beta$ indices will refer to sections of the left-moving bundle $\cV$ and its dual $\cV^\ast$.   We will continute to denote the bosonic coordinates by $y,\yb$.  Thus, $\chi^\alpha$ ($\chib_\alpha$) transforms as a section of the pullback of $\cV$ ($\cV^\ast$).  In particular, the $\chi$ kinetic term is
\begin{align}
2\pi L \supset \chib_\alpha \Dbz \chi^\alpha = \chib_\alpha ( \pbz \chi^\alpha + \pbz y^I \cA^\alpha_{ I\beta} \chi^\beta)~, 
\end{align}
where $\cA$ is a HYM connection on $\cV$ with traceless curvature $\cF= \pb \cA$. 

Using the connection, we can easily describe the full set of operators that can give rise to gauge-netural massless states in the (NS,R) sector (we ignore the universal gravitino and dilatino states and drop the normal ordering):
\begin{align}
\cO^4(z) = \Psi^4_I \p y^I,\qquad
\cO^{5+6}(z) = \Psi^{5\alpha}_{\beta} \chib_\alpha \chi^\beta + \Psi^{6 I} (\rho_I - \cA^\alpha_{I \beta} \chib_\alpha \chi^\beta)~.
\end{align}
As in our discussion of states in the $k=1$ sector we suppressed the expansion of each of these in $\etab$; taking that into account the wavefunctions correspond to the following bundles:
\begin{align}
\Psi^4 \in \Gamma(\oplus_u \Omega^{(0,u)}(T^\ast_B))~,\qquad 
\Psi^5 \in \Gamma(\oplus_u \Omega^{(0,u)}(\End \cV))~,\qquad
\Psi^6 \in \Gamma(\oplus_u \Omega^{(0,u)}(T_B))~.
\end{align}
These states are $\bQb_0$ closed iff $\Psi^4$, $\Psi^5$ and $\Psi^6$ are $\pb$-closed and
\begin{align}
\obs(\Psi^6) + \pb \Psi^5 = 0~,
\end{align}
where $\obs(\Psi^6_u)$ is a (0,u+1) $\pb$-closed form valued in (traceless) endomorphisms of $\cV$
\begin{align}
\obs(\Psi^6)^\alpha_{\beta \Jb_0 \cdots \Jb_u} \equiv \Psi^{6I}_{[\Jb_1 \cdots\Jb_u} \cF^\alpha_{\Jb_0] I\beta}~.
\end{align}
Taking cohomology, $[\obs(\Psi^6_u)] \in H^{u+1}(B,\End \cV)$.  As explained in~\cite{Anderson:2011ty}, at $u=1$ this is the Atiyah class~\cite{Atiyah:1957xx}---an obstruction to extending infinitesimal complex structure deformations of the base $B$ to infinitesimal complex structure deformations of the holomorphic bundle $\cV\to B$.  Thus, our states fit into the complex
\begin{align}
\xymatrix@R=1mm@C=1cm{
\cO^4_0 \ar[r]^-{\bQb_0} &
\cO^4_1  \ar[r]^-{\bQb_0} &\cO^4_2   \ar[r]^-{\bQb_0} &\cO^4_3\\
{\begin{matrix} \cO^5_0 \\ \oplus \\ \cO^6_0 \end{matrix}} \ar[r]^-{\bQb_0}&
{\begin{matrix} \cO^5_1  \\ \oplus\\
\cO^6_1 
\end{matrix}} \ar[r]^-{\bQb_0} & 
{\begin{matrix}
\cO^5_2 \\ \oplus\\
\cO^6_2 
\end{matrix}} \ar[r]^-{\bQb_0} & 
{\begin{matrix} \cO^5_3 \\ \oplus \\ \cO^6_3 \end{matrix}}
}
\end{align}
Taking $\bQb_0$ cohomology we find
\begin{align}
\xymatrix@R=1mm{
0 & H^1(T^\ast) \ar[r]^{0} & H^2(T^\ast) & 0 \\
{\begin{matrix} H^0(\End\cV) \\ \oplus \\H^0(T) \end{matrix}}  \ar[r]^{\obs_0} &
{\begin{matrix} H^1(\End\cV) \\ \oplus \\H^1(T) \end{matrix}}  \ar[r]^{\obs_1} &
{\begin{matrix} H^2(\End\cV) \\ \oplus \\H^2(T) \end{matrix}}  \ar[r]^{\obs_2} &
{\begin{matrix} H^3(\End\cV) \\ \oplus \\H^3(T) \end{matrix}} . 
}
\end{align}
For traceless $\End \cV$ on the CY 3-fold $B$
\begin{align}
H^0(B,\End\cV) = H^3(B,\End\cV)=0~,\qquad H^2(\End\cV) \simeq \overline{H^1(B,\End\cV)}~,
\end{align}
so that the complex reduces to
\begin{align}
\xymatrix{
0 & H^1(T^\ast) \ar[r]^{0}& H^2(T^\ast) & 0 \\
0 &
{\begin{matrix} H^1(\End\cV) \\ \oplus \\H^1(T) \end{matrix}}  \ar[r]^{\obs_1} &
{\begin{matrix} H^2(\End\cV) \\ \oplus \\H^2(T) \end{matrix}}   &
0
}
\end{align}
The only Atiyah obstructions arise in $H^1(B,T) \to H^2(B,\End\cV)$, and hence there are
\begin{align}
h^1(T^\ast) + h^1(T) + h^1(\End\cV) - \dim\ker \obs_1
\end{align}
massless gauge-neutral singlets.

The patient reader who has made it to this last appendix may perhaps be aware that in a (0,2) NLSM with a tree-level $H$-flux there are additional 
obstructions similar to the $H^1(B,T) \to H^2(B,\End\cV)$ map just discussed~\cite{Melnikov:2011ez}.  
The $B$-field coupling will alter the $\eta$ equations of motion and lead to $H$-flux appearing in $\bQb_0 \cdot \rho$, and we expect that including this contribution should reproduce the result of~\cite{Melnikov:2011ez}.  
It would be useful to check that in detail.

\bibliographystyle{./utphys}
\bibliography{./bigref}

\end{document}

%% file: basedefs2.tex
\DeclareFontFamily{U}{rsf}{}
\DeclareFontShape{U}{rsf}{m}{n}{
  <5> <6> rsfs5 <7> <8> <9> rsfs7 <10-> rsfs10}{}
\DeclareMathAlphabet\Scr{U}{rsf}{m}{n}

\def\CO#1#2{{[#1,#2]}}
\def\AC#1#2{{\{#1,#2\}}}

\def\iden{{\mathbbm 1}}

\def\rep#1{{{\boldsymbol{#1}}}}
\def\brep#1{{{\overline{\boldsymbol{#1}}}}}

\def\cDb{{\overline{\cD}}}
\def\cQb{{\overline{\cQ}}}

\def\GUL{\GU(1)_{\text{L}}}
\def\GUR{\GU(1)_{\text{R}}}

\def\C{{\mathbb C}}
\def\H{{\mathbb H}}
\def\Q{{\mathbb Q}}
\def\P{{\mathbb P}}
\def\R{{\mathbb R}}
\def\Z{{\mathbb Z}}

\def\End{\operatorname{End}}

\def\Pic{\operatorname{Pic}}

\def\ch{\operatorname{ch}}

\def\SO{\operatorname{SO}}

\def\SU{\operatorname{SU}}
\def\GU{\operatorname{U{}}}

\def\GE{\operatorname{E}}

\def\so{\operatorname{\mathfrak{so}}}

\def\su{\operatorname{\mathfrak{su}}}
\def\Lu{\operatorname{\mathfrak{u}}}
\def\Le{\operatorname{\mathfrak{e}}}

\def\p{\partial}
\def\pb{\bar{\partial}}

\def\la{\langle}
\def\ra{\rangle}

\def\ff#1#2{{\textstyle\frac{#1}{#2}}}
\def\half{\frac{1}{2}}

\def\cA{{\cal A}}

\def\cC{{\cal C}}
\def\cD{{\cal D}}
\def\cE{{\cal E}}
\def\cF{{\cal F}}
\def\cG{{\cal G}}
\def\cH{{\cal H}}

\def\cJ{{\cal J}}

\def\cL{{\cal L}}
\def\cM{{\cal M}}

\def\cO{{\cal O}}

\def\cQ{{\cal Q}}

\def\cS{{\cal S}}
\def\cT{{\cal T}}
\def\cU{{\cal U}}
\def\cV{{\cal V}}
\def\cW{{\cal W}}
\def\cX{{\cal X}}
\def\cY{{\cal Y}}

\def\ep{{\epsilon}}

\newcommand\alphab{\overline{\alpha}}
\newcommand\betab{\overline{\beta}}
\newcommand\gammab{\overline{\gamma}}

\newcommand\epb{\overline{\ep}}

\newcommand\etab{\overline{\eta}}
\newcommand\thetab{\overline{\theta}}

\newcommand\lambdab{\overline{\lambda}}

\newcommand\xib{\overline{\xi}}

\newcommand\phib{\overline{\phi}}
\newcommand\chib{\overline{\chi}}

\newcommand\nut{\widetilde{\nu}}

\newcommand\Sigmat{\widetilde{\Sigma}}

\newcommand\cb{\overline{c}}

\newcommand\gb{\overline{g}}

\newcommand\ib{\overline{\imath}}
\newcommand\jb{\overline{\jmath}}
\newcommand\kb{\overline{k}}

\newcommand\mb{\overline{m}}

\newcommand\ub{\overline{u}}

\newcommand\xb{\overline{x}}
\newcommand\yb{\overline{y}}
\newcommand\zb{\overline{z}}

\newcommand\htld{\widetilde{h}} 

\newcommand\Vh{\widehat{V}}

\newcommand\Ab{\overline{A}}

\newcommand\Db{\overline{D}}

\newcommand\Hb{\overline{H}}
\newcommand\Ib{\overline{I}}
\newcommand\Jb{\overline{J}}
\newcommand\Kb{\overline{K}}
\newcommand\Lb{\overline{L}}

\newcommand\Pb{\overline{P}}

\newcommand\Tb{\overline{T}}

\newcommand\Vb{\overline{V}}
\newcommand\Wb{\overline{W}}

\newcommand\Yb{\overline{Y}}
